\documentclass{aa}  

\usepackage{graphicx}
\usepackage{txfonts}
\usepackage{upgreek}
\usepackage[colorlinks=true,     linkcolor=blue, citecolor=blue, filecolor=blue, urlcolor=blue]{hyperref}

\begin{document} 

   \title{\texttt{viper}: High-precision radial velocities from the optical to the infrared}

   \subtitle{Reaching 3 m/s in the $K$ band of CRIRES$^{+}$ with telluric modelling}

   \author{J. K\"ohler
          \inst{1}
          \and
          M.~Zechmeister\inst{2}
          \and
          A. Hatzes\inst{1}
          \and
          S. Chamarthi\inst{1}
          \and
          E. Nagel\inst{2}
          \and
          U. Seemann\inst{2,3}
          \and
          P. Ballester\inst{3}
          \and
          P.~Bristow\inst{3}
          \and
          P.~Chaturvedi\inst{4,1}
          \and
          R.~J.~Dorn\inst{3}
          \and
          E.~Guenther\inst{1}
          \and
          V.~D.~Ivanov\inst{3}
          \and
          Y.~Jung\inst{3}
          \and
          O.~Kochukhov\inst{5}
          \and
          T.~Marquart\inst{5}
          \and
          L.~Nortmann\inst{2}
          \and
          R.~Palsa\inst{3}
          \and
          N.~Piskunov\inst{5}
          \and
          A.~Reiners\inst{2}
          \and
          F.~Rodler\inst{6}          
          \and
          J.~V.~Smoker\inst{6}
          }

   \institute{TLS Tautenburg,
              Sternwarte 5, 07778 Tautenburg, Germany \\
              \email{jana@tls-tautenburg.de}
         \and
             Universit\"at G\"ottingen, Institut f\"ur Astrophysik, Friedrich-Hund-Platz 1,
              37077 G\"ottingen, Germany
         \and
             European Southern Observatory, Karl-Schwarzschild-Str. 2, 85748 Garchingen, Germany
         \and
             Department of Astronomy and Astrophysics, Tata Institute of Fundamental Research, 400005 Mumbai, India
         \and
             Department of Physics and Astronomy, Uppsala University, Box 516, 75120 Uppsala, Sweden
         \and
             European Southern Observatory, Alonso de Cordova 3107, Vitacura, Casilla 19001, Santiago, Chile
             }

   \date{Received 27 January 2025 / Accepted 22 April 2025}

  \abstract 
   {High-precision radial velocity (RV) measurements with slit spectrographs require the instrument profile (IP) and Earth's atmospheric spectrum to be known and to be incorporated into the RV calculation.}
   {We developed an RV pipeline, called Velocity and IP EstimatoR ({\tt viper}), to achieve high-precision RVs in the near-infrared (NIR). The code is able to process observations taken with a gas cell and includes modelling of the IP and telluric lines.}
   {We utilised least-square fitting and telluric forward modelling to account for instrument instabilities and atmospheric absorption lines. As part of this process, we demonstrate the creation of telluric-free stellar spectra.}
   {By applying {\tt viper} to observations obtained with the upgraded CRyogenic high-resolution InfraRed Echelle Spectrograph (CRIRES$^{+}$) and a gas absorption cell in the \textit{K}~band, we are able to reach an RV precision of around 3~m/s over a time span of 2.5~years. For observations using telluric lines for the wavelength reference, an RV precision of 10~m/s is achieved.}
   {We demonstrate that despite telluric contamination, a high RV precision is possible at NIR wavelengths, even for a slit spectrograph with varying IP. Furthermore, we show that CRIRES$^{+}$ performs well and is an excellent choice for science studies requiring precise stellar RV measurements in the infrared.}


   \maketitle

\section{Introduction}

In recent years, a number of new instruments and data reduction pipelines have been developed to obtain high-precision radial velocities (RVs).
In particular in the optical, considerable progress has been made and RV precision below 50~cm/s has been reached, for example by ESPRESSO and MAROON-X \citep[e.g.][]{espresso, basant}. Yet, the RV precision in the near-infrared (NIR) is trailing behind. This is due to a number of factors, such as imprinted atmospheric absorption lines, lower stellar information content, different types of detectors, and usable calibration lamps.
However, observations in the NIR are important for the search and study of exoplanets around cool low-mass stars that are faint at optical wavelengths. Not only are M~dwarfs brightest in the NIR, the signal of stellar activity is also reduced at longer wavelengths.
For the detection of super-Earths around these stars, an RV accuracy of better than 3~m/s is required.

Only recently has an RV accuracy of less than 1~m/s been achieved, with NIRPS in the \textit{H}~band \citep{nirps}.
In the \textit{K}~band, an accuracy of 5~m/s over a timescale of one year was reached with the iSHELL spectrograph using a methane gas cell \citep{ishell_rv}. 
With its ammonia gas cell, the original CRyogenic high-resolution
InfraRed Echelle Spectrograph (CRIRES) achieved a long-term RV accuracy of 5.4~m/s in the \textit{K}~band \citep{bean}.
Whereas a line-by-line approach was used on telluric-corrected spectra to determine the NIRPS RVs, the results for iSHELL and CRIRES were obtained using a forward model that included parameters for the telluric spectrum.
Here, a double challenge arises from the fact that CRIRES and iSHELL are not fibre-fed (as is the case with high-precision spectrographs such as NIRPS) which results in a varying instrument profile (IP). Therefore, the tellurics vary not only in depth but also with the changing IP.
We face the same challenge in our software.

In this paper we introduce the RV pipeline {\tt viper}\footnote{\url{https://mzechmeister.github.io/viper_RV_pipeline}} \citep[Velocity and IP EstimatoR;][]{viper}.
The philosophy of {\tt viper} is to offer a publicly available and user-friendly code that is able to process data from various spectrographs. {\tt viper} is officially provided by the consortium that built the upgraded CRIRES \citep[CRIRES$^{+}$;][]{dorn}. 
Originally designed to handle data from optical instruments, following the procedure from \citet{butler}, the code now has been extended to enable the processing of NIR data. {\tt viper} uses a least-square fitting to model the stellar RV as well as the temporal and spatial variable IP. We have improved upon this method by adding a term for the telluric spectrum that enables the forward modelling of molecules present in the Earth's atmosphere. 

In this paper we use CRIRES$^{+}$ observations in the \textit{K}~band to demonstrate {\tt viper}'s ability to handle data in the NIR.
First, we provide an overview of the techniques and algorithms used in {\tt viper}. We show that it is possible to achieve an RV accuracy of 3~m/s over a period of 2.5 years with the use of a gas cell.
Additionally, we present a study of the stability of atmospheric lines in the NIR.
With {\tt viper} it is possible to handle data taken with or without a gas cell, and we show that a long-term RV precision of around 10~m/s can be achieved when using only telluric lines for the wavelength calibration.

\section{The algorithm}
\label{sec:model}

The main concept of {\tt viper} is based on the forward model described in \citet{butler}. 
This involves creating a model that is optimised to best match the observation.
The observations are expected to be performed with the use of a gas cell in the light path, resulting in a spectrum that is a product of stellar and gas cell lines.
An observation of the star without gas cell lines, which we call the stellar template $S_\mathrm{star}$, serves as a reference to calculate differential Doppler shifts $v_\mathrm{star}$ of the stellar absorption lines.
A further discussion on the stellar template will follow in Sect.~\ref{sec:tpl_creation}.
The modelling process requires an ultra-high resolution (spectral resolving power of $\approx$ 10$^6$) transmission spectrum of the gas cell $T_\mathrm{cell}$ that is often obtained with a Fourier transform spectrometer (FTS).
The spectral lines of the absorbing gas in the cell serve as the wavelength reference and enable us to monitor the IP.
For the wavelength solution $\lambda(x)$ at a given pixel $x$, we used a low degree polynomial, using the wavelength solution generated by a data reduction pipeline as an initial guess.
This results in the following forward model ${S(x)}$ for the observed spectrum:
\begin{equation}
\label{eq_butler}
S(x)\, {=} \, k \cdot [ T_\mathrm{cell} \, (\lambda) \cdot S_\mathrm{star}(\lambda( v_\mathrm{star}))] \otimes \mathrm{IP}(\lambda, x),
\end{equation}
\noindent
where $\otimes$ represents the convolution.
The term $k$ is a flux normalisation, which can be just a scalar to handle different flux levels, or a smooth function (low order polynomial) over pixel (or wavelength) to take into account gradients due to imperfect atmospheric dispersion corrections or even the shape of the blaze function.

Although we denote wavelengths, we actually worked with \mbox{logarithmic} wavelengths. This parametrisation is generally more suitable, since the resolution in this domain is usually less variable than in wavelength or pixel space.
Likewise the IP is defined in velocity, which also corresponds to logarithmic wavelengths. 
The over-sampled convolved forward model is still on a discrete, uniform log-wavelength grid. It is linearly interpolated to compute the model at the observed pixels, whose wavelengths is evaluated from the modelled wavelength solution.

Using the least-square fitting algorithm {\tt curvefit} from {\tt scipy.optimize} \citep{scipy}, the parameters for the normalisation, wavelength solution, IP, and Doppler shift are optimised concurrently.
Instrumental shifts, caused by, for example, temperature or pressure fluctuations, are captured by the simultaneous wavelength solution.

At optical wavelengths, iodine is the most common molecule used in absorption cells, as it is rich in sharp lines and enables precise RV measurements at low costs.
Since iodine has very weak absorption lines above $\sim$6\,000~\r{A}, other molecules are required for instruments operating in the NIR range.
In the case of CRIRES$^{+}$, the gas cell is filled with a multi-species gas, containing NH$_3$, $^{13}$CH$_4$, and C$_2$H$_2$ \citep{seemann, dorn}. The Earth’s atmosphere also provides lines usable for the wavelength calibration, although these lines vary strongly in depth and slightly in position, depending on the time of observation (see Sect.~\ref{sec:model_tell}). 

Observations are taken in the Earth rest frame, and therefore, a correction for the barycentric motion is needed. This velocity shift can be written as
\begin{equation}
\label{eq_redb}
z_0 = \frac{\lambda_0}{\lambda'}-1,
\end{equation}
\noindent
with $\lambda'$ being the observed wavelength and $\lambda_0$  the barycentric-corrected wavelength.
For an observer in the barycentre, the measured velocity shift of the star, with respect to the wavelength of the stellar template $\lambda_\mathrm{tpl}$, is 
\begin{equation}
\label{eq_reds}
z = \frac{\lambda_0}{\lambda_\mathrm{tpl}}-1.
\end{equation}
\noindent
The stellar motion, or RV, is $v = cz$, with $c$ being the speed of light.
With $v$ as the input parameter we are interested in, the wavelengths is Doppler shifted according to
\begin{equation}
\label{eq_red}
\lambda_\mathrm{tpl} =  \lambda' \frac{1+z_0} {1+ \frac{v}{c}}.
\end{equation}
\noindent
We chose in {\tt viper} the optimisation with non-relativistic velocities, which is a fair assumption for our science purposes. Yet, we applied the barycentric correction to the stellar template beforehand using relativistic velocities.
A more detailed description can be found in \citet{berv}. For the calculation of the barycentric correction, we used the {\tt SkyCoord.radial\_velocity\_correction} function from {\tt astropy} \citep{astropy}.

As the model parameters vary with wavelength, the spectrum is processed in wavelength segments, or chunks, which can be as large as one echelle order. 
For each chunk indexed with $o$, the least-square fitting provides a covariance matrix, whose diagonal provides formal uncertainties for each parameter. For the RVs, we use $\epsilon_o$ to denote the uncertainty.
The final RV is a weighted mean of the RVs calculated for each individual chunk: 
\begin{equation}
\label{eq_RV}
\mathrm{RV} = \frac{\sum \epsilon_o^{-2} \cdot \mathrm{RV}_{o} } {\sum \epsilon_o^{-2}} .
\end{equation}
The corresponding uncertainty is calculated as\begin{equation}
\label{eq_eRV}
\epsilon_{\mathrm{RV}} = \sqrt{ \frac{1}{N_o-1} \cdot \frac{\sum \epsilon_o^{-2} \cdot (\mathrm{RV}_o -\mathrm{RV})^{2}}{\sum \epsilon_o^{-2}} },
\end{equation}
with $N_{o}$ being the number of processed chunks.

\subsection{Telluric forward modelling}
\label{sec:model_tell}

Molecules present in our Earth's atmosphere lead to emission and absorption features in the observed spectra.
Masking out the telluric affected regions is a common approach in the optical region where the tellurics line density is low. This is unfeasible in the NIR where the line density is increasing dramatically and nearly the entire wavelength range is contaminated. 
Indeed, masking telluric features with a depth greater than 2\% eliminates up to 50\% of the spectrum in the \textit{K}~band. Yet, micro-tellurics (with depths lower than 2\%) do have an impact on the accuracy of the measured RVs.

One common way to deal with this problem is to use synthetic spectra of the Earth's atmosphere to model and remove telluric lines from the observed spectra. Prominent codes for this purpose are {\tt MolecFit} \citep{smette}, {\tt TelFit} \citep{telfit}, and {\tt Tapas} \citep{tapas}. 
Although it is possible to run {\tt viper} on spectra corrected in advance for tellurics, we chose to model the tellurics simultaneously.
The idea of using telluric lines for the wavelength calibration was first proposed by \citet{griffin}. 
There also have been a few attempts to include tellurics in a forward modelling process. This was done, for example, by \citet{bean} for the RV determination of the original CRIRES and more recently by \citet{subaru} for the data of the IRD spectrograph and by \citet{ishell_rv} for the data of the iSHELL spectrograph.

By extending Eq.~\eqref{eq_butler} with a term for the telluric absorption lines, the forward modelling becomes
\begin{equation}
\label{eq_butler_atm}
S(x)\, {=} \, k \cdot [ T_\mathrm{cell} \, (\lambda) \cdot T_\mathrm{atm}\, (\lambda, v_\mathrm{atm}) \cdot S_\mathrm{star}  (\lambda, v_\mathrm{star})] \otimes \mathrm{IP}(\lambda, x).
\end{equation}
\noindent
Here $T_\mathrm{atm}$ represents the telluric model, which is a product of the transmission of individual molecules:
\begin{equation}
\label{eq_prod_atm}
T_\mathrm{atm} \, {=} \, \prod_\mathrm{molec} T_\mathrm{molec} (\lambda, \alpha_\mathrm{molec}).
\end{equation}
\noindent

\begin{figure}
\centering
\includegraphics[ width=0.49\textwidth]{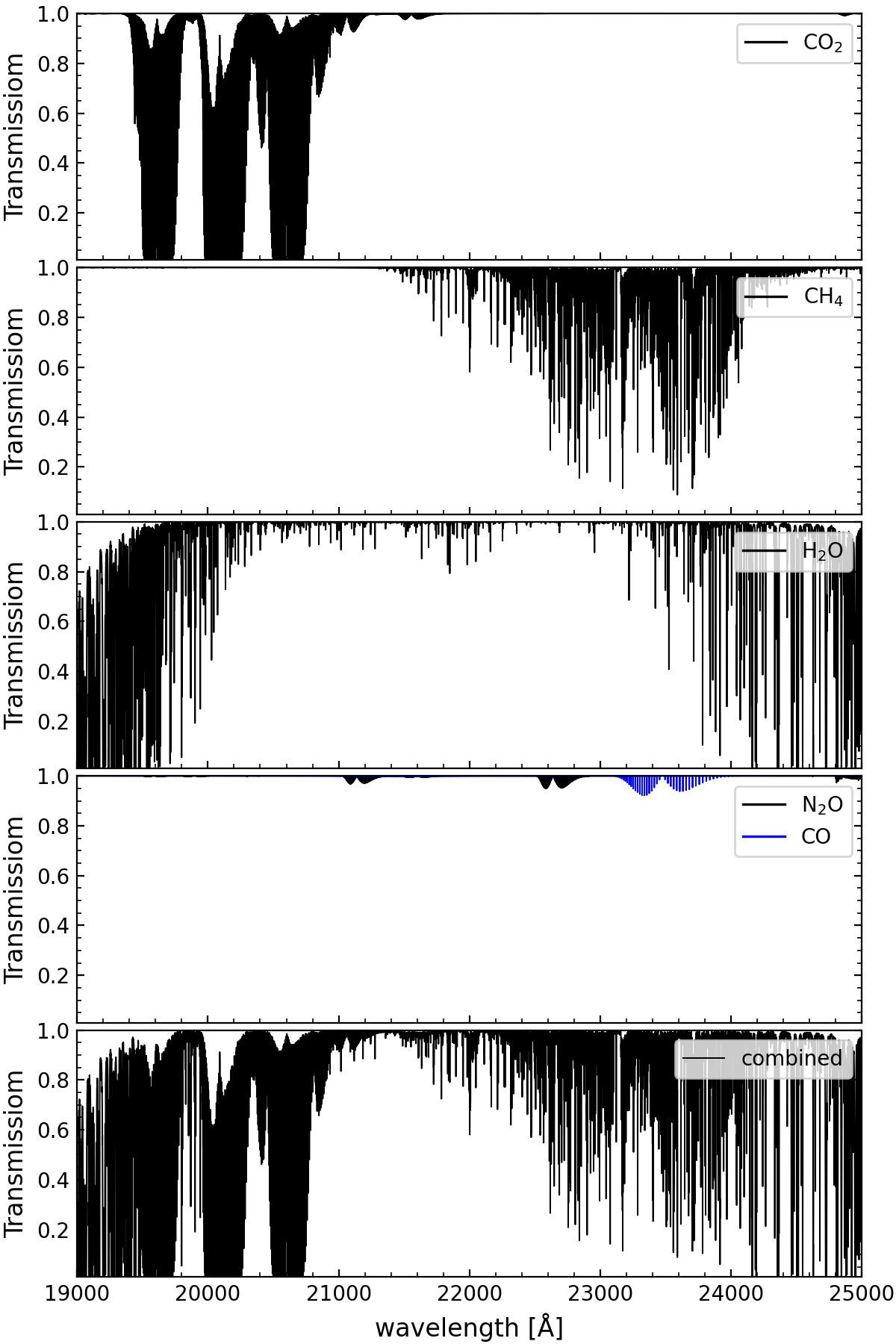}
\caption{Synthetic spectra in the \textit{K}~band for molecules present in the Earth's atmosphere, calculated for an airmass of 1, and an IWV of 1~mm for water. The first, second, and third panels show the spectra of the dominant species CO$_2$, CH$_4$, and H$_2$O, while the fourth panel shows the less dominant contributors N$_2$O and CO. The bottom panel shows the combined spectrum. }
\label{fig:spec_tell}
\end{figure}

In the \textit{K}~band, the dominating molecules are water, methane, and carbon dioxide (Fig.~\ref{fig:spec_tell}). 
Their line depths depend on the airmass and the molecular column density, which vary over time. In particular, the water content can vary a lot, which is described by the integrated water vapour (IWV). 
The IWV is calculated via an integral over a density function of the condensed water through the entire atmosphere in the line of sight.
The radiative transfer equation for the telluric transmission spectrum is 
\begin{equation}
\label{eq_rad_trans}
T_\mathrm{molec}(\lambda) \, {=} \, e ^{- \sigma_\mathrm{molec}(\lambda) \cdot n_\mathrm{molec} \cdot z},
\end{equation}
\noindent
where $\sigma$ is the effective cross-section, $n$ the average number density and $z$ the path length (related to airmass). For water, the average number density is related to the IWV. 
The changing line depth therefore can be adjusted using an exponent: 
\begin{equation}
\label{eq_atm_scal}
 T_\mathrm{molec} \, (\lambda) \, {=} \, T_\mathrm{molec, syn} \, (\lambda) ^ \mathrm{\alpha_\mathrm{molec}} .
\end{equation} 
\noindent
Since airmass and density are degenerate, one model parameter is sufficient.
For ${\alpha_\mathrm{molec}}$\,=\,1, we have the components of our reference atmosphere, which was computed for airmass\,=\,1 and for IWV\,=\,1~mm.
Thus, the parameter ${\alpha_\mathrm{molec}}$ corresponds to airmass and, in the case of water, to IWV.
A detailed study can be found in \citet{telluric}. 

The synthetic atmosphere model $T_\mathrm{molec, syn}(\lambda)$, which we used as a starting guess, was created for each molecule using a line-by-line radiative transfer model
\citep[{\tt LBLRTM};][]{LBLRTM1, LBLRTM2} and characteristic parameters for the location of the Very Large Telescope (VLT).
During the optimisation procedure, the exponent ${\alpha_\mathrm{molec}}$ will be fitted for each molecule to correct for changes caused by airmass, IWV and seasonal effects. 

Another option would be the application of only two coefficients -- one for water and a combined one for all other molecules.
This is possible under the assumption that seasonal variations of the non-water molecules are expected to be small compared to the influence of a varying airmass.
However, testing demonstrated that in most cases a better RV precision was achieved by the usage of a individual coefficient for each molecule. 
For observations with a signal-to-noise ratio (S/N) below 50, we cannot distinguish between micro-tellurics and noise. With the assignment of a combined coefficient for non-water molecules, the modelling of molecules with weaker absorption (e.g. N$_2$O and CO) will benefit from a correct modelling of molecules with stronger absorption  (e.g. CH$_4$ and CO$_2$).
Within {\tt viper}, both options for the coefficients are available.

Supplementary to the exponents, an additional parameter $v_\mathrm{atm}$ for the telluric Doppler shift is introduced. From the study by \citet{figueira}, it is known that the telluric lines can be stable down to 10~m/s depending on changes in wind speed and the velocity of atmospheric components such as the jet stream. 
Although incorrect telluric Doppler shifts have less of an impact when additional gas cell lines are used for the wavelength calibration, a correction of these shifts is required when aiming for 3~m/s or below.
The current implementation of the code applies a common shift for all molecules, assuming all molecules experience the same weather conditions.
This is obviously not entirely true, as the variations of the volume-mixing ratio with altitude are different for the various molecules.

Currently, our model is not able to adjust for changing line profiles, as caused by variations of the air pressure \citep[see e.g.][]{smette, allart}. Under the premise of small pressure variations, the profile is expected to remain stable and changes to be negligible compared to instrumental effects. 
This is a valid assumption for the VLT at Paranal, where we have calculated a mean air pressure of 744.35~hPa with a standard deviation of 1.31~hPa for the 2.5 years of our observations. 
As a side note, for the temperature we calculate a mean value of 12.16$^{\circ}$C with a standard deviation of 2.53$^{\circ}$C. 
It should be noted that these values represent the conditions on the ground and are expected to vary in the higher layers of the atmosphere.

\citet{wang} simulated and discussed the impact of a wrong telluric line profile. For $K$-band data, the telluric forward modelling with an incorrect telluric line profile results in an RV scatter of 6.8~m/s. For other discussed bands, which have less strong telluric lines, this value varies between 0.008~m/s (\textit{B}~band) and 3.5~m/s (\textit{J}~band). By their definition of the model, the RV scatter would be 0~m/s under the assumption of a correct telluric line profile. As there is no explicit declaration of the deviation from the correct line profile, the obtained values should be considered an upper limit.
 
\begin{figure*}
\centering
\includegraphics[ width=0.99\textwidth]{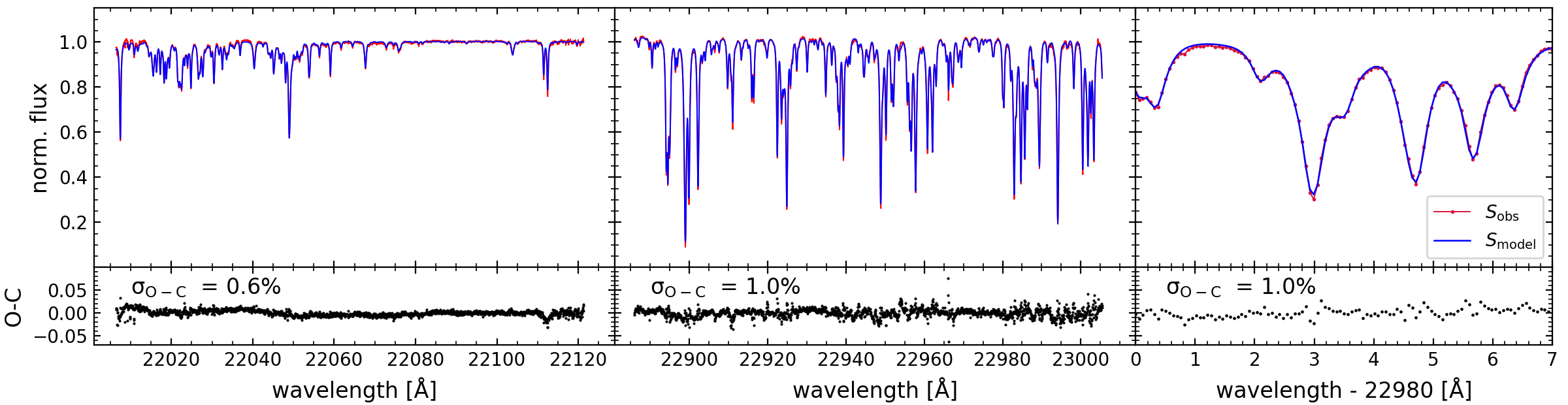}
\caption{CRIRES$^{+}$ observation of the telluric standard star ${\alpha}$ Aql from 17 August 2021 (airmass = 1.47, IWV = 3.22~mm) for two different chunks (left and center panels) and a zoom-in (right panel). Top panels: Observed spectrum (red) without the gas cell, and the best-fit telluric model (blue) from {\tt viper}. Bottom panels: Residuals and their standard deviations.}
\label{fig:Tell_corr}
\end{figure*}

For a first test, we used telluric standard stars, which are hot (i.e. with few lines), rapidly rotating (shallow and broad lines) stars whose observed spectra are dominated by the telluric features.
Considering the example of ${\alpha}$ Aql (A7 star, m$_K$~=~0.24~mag), the telluric forward modelling with {\tt viper} is demonstrated in Fig.~\ref{fig:Tell_corr} for two parts of the spectrum. The observation was taken at an airmass of 1.47 and an IWV of 3.22~mm, the S/N is about 300. 
For the spectral order containing telluric lines with transmission depths of 0.4, the standard deviation of the residuals is 0.6\%. 
For the spectral order contaminated by strong tellurics, down to a transmission value of 0.1, the standard deviation of the residuals is 1\%, leaving visible structures in the residuals.
Observations of the same star with lower airmass and IWV, are modelled accurately to 0.5\% and 0.75\% for the same wavelength regions (not plotted here). For demonstration purposes an observation affected by deep telluric lines was chosen to demonstrate the performance capability of {\tt viper}. 
It should be noted that the precise modelling of deep lines is a common problem in all telluric correction codes. 
Thus, our results are in good agreement with the residual values between 0.6\% and 1.2\% as reported by \citet{smette}, when running {\tt MolecFit} on $K$-band data from the original CRIRES instrument.

\begin{figure*}
\centering
\includegraphics[ width=0.99\textwidth]{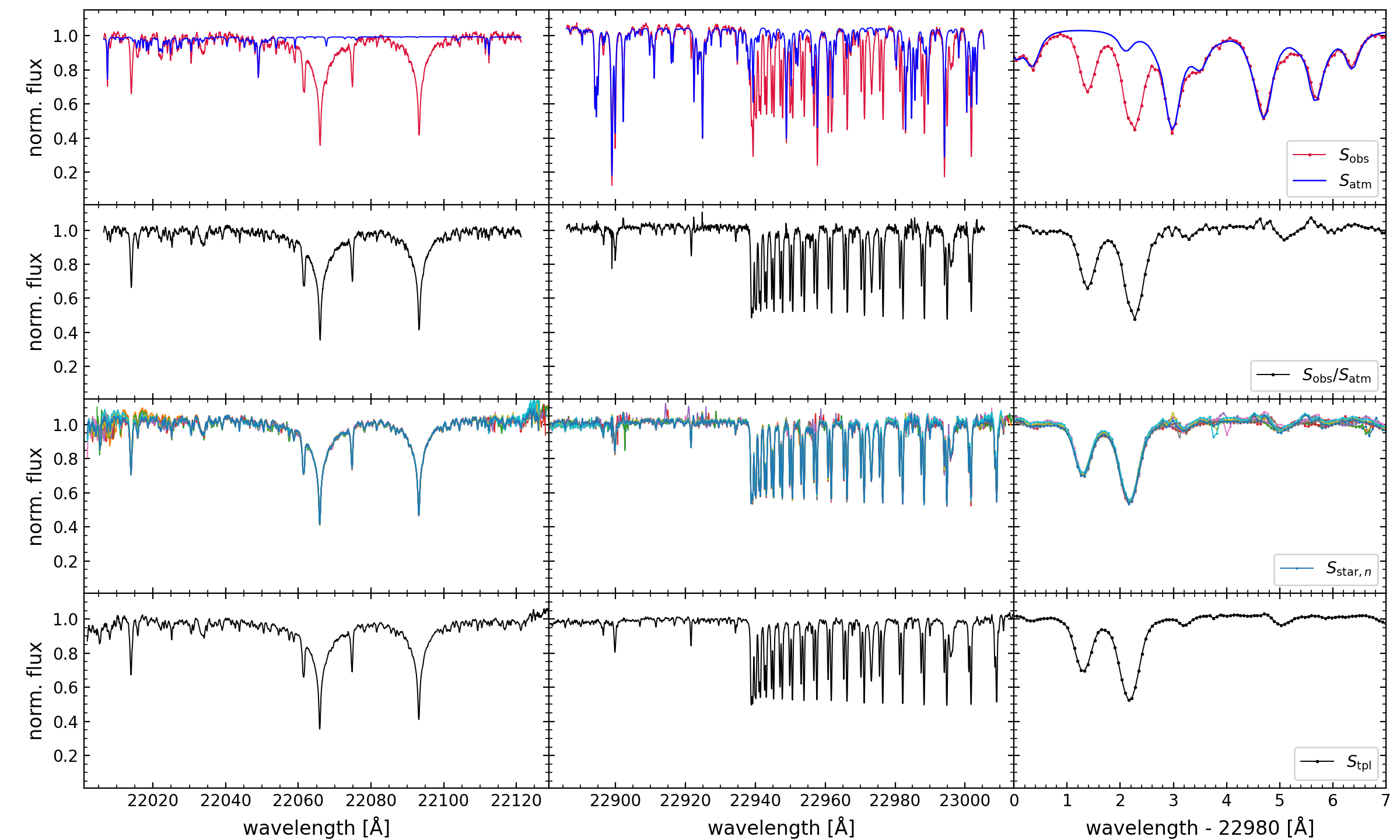}
\caption{Telluric correction and template generation of the M~dwarf GJ\,588 for two different chunks. First row: CRIRES$^{+}$ observation from 8 August 2022 of the star (red) overlaid with the best-fit telluric model obtained with {\tt viper} (blue). Second row: Telluric-corrected spectrum. Third row: Individual telluric-corrected spectra taken between March and August 2022. Last row: Co-added spectrum using the average of the individual observations from the plot above. This represents the final stellar template, which is used as reference for the RV determination.}
\label{fig:Tell_corr_star}
\end{figure*}

\begin{figure*}
\centering
\includegraphics[ width=0.99\textwidth]{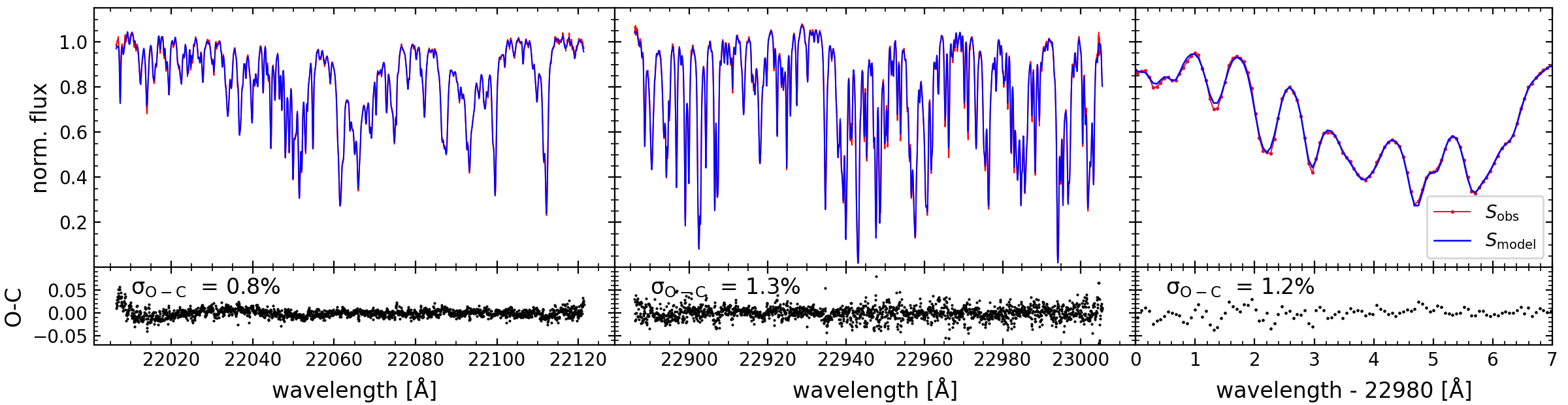}
\caption{Same as Fig. \ref{fig:Tell_corr} but for the M~dwarf GJ\,588 observation (obtained with the gas cell) from 8 August 2022 (airmass = 1.3, IWV = 1.23~mm). 
}
\label{fig:model_gj588}
\end{figure*}

\subsection{Creation of telluric-free stellar templates}
\label{sec:tpl_creation}

In this section we discuss the removal of telluric lines from observations of M~dwarfs. 
This is necessary for the creation of the telluric-free stellar templates, which later will serve as reference for the RV determination. 
It is recommended to use a template generated from stellar observations over the use of a synthetic spectrum, for example that obtained from PHOENIX \citep{phoenix}. 
As discussed in earlier papers, observed stellar templates produce better RV results \citep[e.g. ][]{figueira_cr}. Our tests confirmed this. 

{\tt viper} offers the option of generating a stellar template by co-adding several observations.
In a first step, the telluric lines are modelled and removed from the individual spectra, which were obtained without a gas cell.
We adjusted Eq.~\eqref{eq_butler_atm} by setting $T_\mathrm{cell}$ and $S_\mathrm{star}$ to one to derive the telluric model:
\begin{equation}
\label{eq_butler_nostel}
S_\mathrm{atm}(x) \, {=} \, k \cdot \, T_\mathrm{atm}\, (\lambda) \otimes \mathrm{IP(\lambda, x)}.
\end{equation}
\noindent
After optimising all model parameters, we divided our observation by $S_\mathrm{atm}$ to get the telluric-free stellar spectrum: 
\begin{equation}
\label{eq_butler_notell}
S_\mathrm{star}(\lambda) \, {=} \, \frac{S_\mathrm{obs}(\lambda)}{ S_\mathrm{atm}\, (\lambda)}
.\end{equation}
\noindent 
It should be noted that the deconvolution of the spectra is not yet implemented in the template creation process with {\tt viper}. 

The imperfections of the atmospheric model becomes visible in the residuals of the corrected spectrum at the locations where particularly deep telluric lines occur.
To address this issue and further increase the S/N of the stellar template, it is recommended to combine observations from different epochs. 
Telluric features are relatively stable in wavelength with respect to stellar lines, but unlike the latter are not subjected to the barycentric motion of the Earth. Co-adding several spectra that were taken at different barycentric velocities, and are corrected for that, helps reduce residuals from the telluric correction, as they are averaged out.

To ensure a good quality of the final telluric-free stellar template $S_\mathrm{tpl}$, a weighted mean was applied:
\begin{equation}
\label{eq_tpl}
S_\mathrm{tpl}(\lambda) = \frac{\sum w_n(\lambda) \cdot S_{\mathrm{star}, n}(\lambda) } {\sum w_n(\lambda)},
\end{equation}
\noindent
with $n$ denoting the observations at different epochs.
The weighting $w_n(\lambda)$ decreases linearly with the telluric transmission at a given wavelength, as residuals are expected to be stronger for deeper lines.
Additionally, noise contaminated parts, represented by the flux error ${\epsilon_{S_{\mathrm{star},n}}}$, were down-weighted.
The final weighting factor becomes
\begin{equation}
\label{eq_tpl_wei}
w_{n}(\lambda) = \frac{T_{\mathrm{atm},n}(\lambda) } { \epsilon_{S_{\mathrm{star},n}}(\lambda) ^{2} }.
\end{equation}

Figure~\ref{fig:Tell_corr_star} illustrates this procedure on the M~dwarf GJ\,588. The top panels demonstrate the telluric correction applied on a single observation in two spectral orders. The corresponding corrected spectrum is plotted in the second row. In the third row, 12 observations, taken between March and August 2022, are plotted above each other, after correcting for the barycentric motion. Although all output stellar spectra are in good agreement with each other, systematics in the residuals become visible especially for the order contaminated by deep tellurics. These disappear in the co-added spectrum, plotted in the bottom panels.
For the later RV calculation, the template will be trimmed at the edges, to get rid of wavelength ranges with low S/N.

The method is straightforward for RV standard stars (stars that show moderate RV variability).
However, the situation becomes more complex when Doppler shifts are present, such as in the case of a star with an existing exoplanet. In such cases, an alternative approach is required. In a first step, we applied the telluric correction to one observation, preferably with a high S/N. Afterwards, this spectrum can serve as reference template for all of the other observations. This is done by using Eq.~\eqref{eq_butler_atm} and setting $T_\mathrm{cell}$ to unity.
Before the co-adding, the wavelengths will be corrected for the calculated RV shifts.
To improve the quality of the final stellar template, this procedure is repeated, taking the recent generated spectrum as template input for another iteration.

In Fig.~\ref{fig:model_gj588}, the created stellar spectrum is used as template for the modelling of an observation of GJ\,588 obtained with the gas cell (S/N~$\approx$~350). The model is according to Eq.~\eqref{eq_butler_atm} a product of stellar, cell, and telluric lines. 
The standard deviations of the residuals are 0.8\% and 1.3\% for this M dwarf, and therefore slightly worse compared to the modelling of the early-type telluric standard standard star.

\subsection{Advantages of the telluric forward modelling}
\label{sec:adv_forward}

The decision to choose telluric forward modelling over other techniques comes with a number of advantages. To begin with, using synthetic molecular spectra for the correction of telluric lines saves valuable observing time that would have been spent
to obtain a high signal-to-noise spectrum on a telluric standard.

However, it still has to be clarified whether the forward modelling is preferred over the removal of tellurics before the determination of the RVs. Here, a number of benefits can be given. First, by using a precalculated telluric standard model, no additional code for the telluric correction is needed, which reduces working steps and computing time.
For instance, while the telluric correction with {\tt MolecFit} takes several minutes for one CRIRES$^{+}$ spectrum, the forward modelling in {\tt viper} requires a few seconds for the same dataset.
Second, in regions with a low gas cell line density, telluric lines can serve as an additional wavelength reference, improving the achieved precision. A third benefit comes from the mathematical treatment of the convolution. As $(S_\mathrm{star} \cdot T_\mathrm{atm}) \otimes \mathrm{IP} \neq (S_\mathrm{star} \otimes \mathrm{IP}) \cdot (T_\mathrm{atm} \otimes \mathrm{IP})$, the division of a telluric spectrum is mathematically incorrect and will result in inaccuracies in the obtained stellar spectrum \citep[see ][]{nagel}. Instead, the forward modelling follows the correct order of multiplication and convolution, avoiding this problem. 
However, this would require that we have access to the real stellar spectrum, $S_\mathrm{star}$, which is not the case since we did not follow the correct order of multiplication and convolution during the telluric correction (Eq.~\eqref{eq_butler_notell}).
This results in errors, as shown in \citet{wang} and \citet{nagel}, which are largest in regions with deep tellurics.
We tried to minimise these residuals in the template generation by down-weighting with the telluric transmission depth (Eq.~\eqref{eq_tpl_wei}). 
By co-adding several spectra from different epochs, these residuals should be largely, but not completely, averaged out.

Further, it should be mentioned that various studies have shown a higher RV precision is obtained when using forward modelling instead of the cross-correlation-function method \citep[e.g.][]{wang}.
However, the simultaneous absorption method requires the presence of a cell or at
least telluric lines for the wavelength calibration. For wavelength regions that are free from these lines, no accurate RV calculation is possible. In addition, a stellar template with high S/N that is free of cell and telluric lines is needed. The accuracy of the RVs depends on the quality of the template.

\begin{figure*}
\includegraphics[ width=0.99\textwidth]{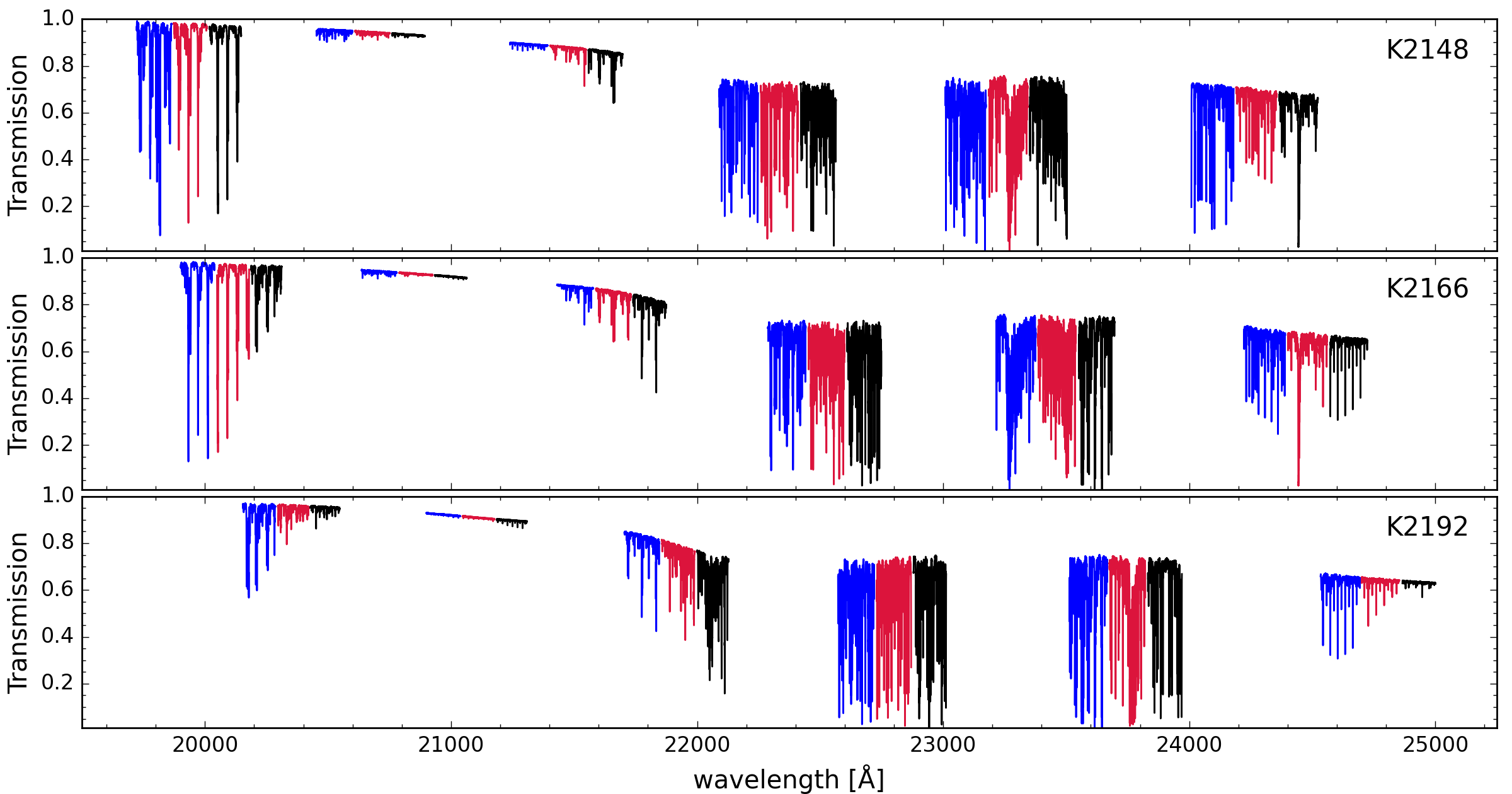}
\caption{Spectrum of the short gas cell from CRIRES$^{+}$ for three settings in the $K$~band, showing the covered wavelength range. The colours mark the three different detectors.}
\label{fig:set_cell}
\end{figure*}

\section{Data processing with {\tt viper}}
\label{sec:dataflow}

{\tt viper} processes 1D spectra as produced by most data reduction pipelines. 
From the input data, a first guess for all parameters is created. These parameters will later be optimised during the least-square fitting. 

As the parameters change with wavelength, and partly even over a small wavelength range, each spectral order should be treated separately. If one order extends over multiple detectors, as for CRIRES$^{+}$, the RV is computed separately for each suborder that falls on one detector.
It is possible to subdivide an order into smaller chunks, of several hundred pixels, to handle changes in the IP along a spectral order. In the case of our observations taken with CRIRES$^{+}$, this did not improve the results. Likely, the IP does not change a lot over an order. Further, smaller chunks can result in a loss of lines for the wavelength calibration. Long chunks are especially helpful for regions with a low line density and observations with low S/N.

We developed {\tt viper} to calculate RVs using various instruments, each with distinct properties.
The user can select various parameter settings such as the polynomial degree for the wavelength solution and normalisation, as well as the IP model, which should be applied. {\tt viper} provides a range of models, from Gaussian or Lorentzian profiles to more complex asymmetric shapes. In addition, it is possible to mask wavelength regions or set thresholds for masking outlier data points, which are, for example, caused by cosmic ray hits.

The current release of {\tt viper} (version~1.1) comes with standard molecular models for the Earth's atmosphere for wavelengths from 3\,500--25\,000~\r{A}, covering all bands from the optical to the NIR. 
For the RV determination it is possible to use just cell lines, just telluric lines, or a combination of the two as the wavelength reference.

\section{Observations}
\label{sec:data}

We used observations obtained with CRIRES$^{+}$ in the \textit{K}~band to demonstrate the capabilities of {\tt viper}.
These data are well suited to test our method and prove that high-precision RV measurements in the reddest part of the NIR are possible.

\subsection{CRIRES$^{+}$}

CRIRES$^{+}$ \citep{dorn} is installed at the ESO VLT in Paranal and operates from 0.95$-$5.3 \textmu m, covering the $YJHKLM$ bands, with a resolving power of around 50\,000 (0.4\arcsec slit) or 100\,000 (0.2\arcsec slit). A gas cell for simultaneous wavelength calibration is available for the $H$ and \textit{K}~band \citep{seemann}. By making use of the gas cell, CRIRES$^{+}$ is expected to reach an RV precision of around 3~m/s. The absorption gas cell is filled with multi-species gas, including NH$_3$, $^{13}$CH$_4$, and C$_2$H$_2$ \citep[resolving power FTS~=~1\,538\,000;][]{seemann}. 
Observations indicate that CRIRES$^{+}$ suffers from instrumental drifts up to 200~m/s within a few hours, and up to 800~m/s ($\sim$1~pixel) on longer timescales, as documented in the CRIRES$^{+}$ User Manual\footnote{\url{https://www.eso.org/sci/facilities/paranal/instruments/crires/doc.html}}.
Without the use of a simultaneous wavelength calibration, either by a gas cell or telluric lines, this would be the limiting RV precision. 

For each of the bands, multiple settings are available, covering different wavelength ranges. The cell spectra for three settings in \textit{K}~band (K2148, K2166 and K2192) are plotted in Fig.~\ref{fig:set_cell}. 
Every setting contains six echelle orders, each falling on the three detectors.

\subsection{Sample selection}

To monitor the stability of the CRIRES$^{+}$ instrument, we selected four RV standard stars from the publicly available RV database {\tt HARPS-RVBANK} published by \citet{trifonov}. We specifically targeted bright M~dwarfs and employed an error-weighted root-mean-square (rms) calculation of the HARPS RVs to identify targets with low RV scatter. \mbox{Table}~\ref{tab_sources} lists our targets together with the number of HARPS spectra ($N_{\mathrm{HARPS}}$) and the derived rms values (rms$_{\mathrm{HARPS}}$).
A direct comparison of our results with HARPS is difficult, as these measurements are made in the optical. In addition, CRIRES$^{+}$ is an instrument with many moving parts and not fibre fed, unlike HARPS.

GJ\,447 has a confirmed planet, but the orbital RV amplitude is smaller than 1.5~m/s and therefore below the RV precision expected for CRIRES$^{+}$. The contribution of the planet signal to our obtained RV results will be discussed in Sect.~\ref{sec:results}.

\subsection{Observation set-up}

Most of our observations were performed with the adaptive optics system and the K2192 setting using the 0.2\arcsec~wide slit ($R\sim$~100\,000).
The exposure time, which is set by the detector integration time (DIT) and the number of such integrations (NDIT), varies for the different stars and even for observations of the same star. Our observations were performed with DIT\,x\,NDIT between 40~s and 75~s. On top of that, several nodding observations were taken to increase the S/N. The purpose of the nodding procedure is to enable the subtraction of the instrument and sky background from the data by placing the star at two different slit positions (here named A and B). For our final spectra, the number of combined noddings varies between 2 and 8 (e.g. AB and AAAABBBB pattern). These integration times yield S/N values between 100 and 350 for our sample of bright stars.

\subsection{Observations}

Many observations, within a time range of a week, were taken during the commissioning runs in June 2021 (COMM3) and August 2021 (COMM4).
Since April 2022, roughly one observation per week of an RV standard star has been taken in the wavelength setting K2192.
In most cases, one spectrum was taken with the gas cell and one without it. 
This gives us the opportunity to study the stability of the instrument over a timescale of more than two and a half years. 
Furthermore, during the commissioning runs in 2021, two time series of GJ\,588 and GJ\,784 over a time span of around 3.7~hours were taken. This allows us to study the stability on short timescales. The observations taken without the gas cell are combined to create a high S/N template, while the observations with gas cell are used to calculate the RVs.
The results of the first two commissioning runs, which were performed in January 2021 (COMM1) and February 2021 (COMM2), were omitted from this analysis due to technical problems with the instrument during that time.

\subsection{Data reduction}

The raw data were pre-processed with the ESO data reduction system (DRS) {\tt cr2res}\footnote{\url{https://www.eso.org/sci/software/pipelines/cr2res/cr2res-pipe-recipes.html}} (version~1.3.0).
First, we reduced dark, flat field, and wavelength calibration frames following the reduction steps described in the {\tt cr2res} manual.
Afterwards, the science observations were reduced with the {\tt cr2res\_obs\_nodding} recipe, taking the produced calibration files as additional input. 
The spectra from the A and B nodding positions are co-added to a final spectrum, using the DRS pipeline.
The blaze correction was done with the help of a 1D spectrum of a flat field observation.

For the RV calculation with {\tt viper}, we chose the wavelength region between 21\,300 and 24\,700~\r{A}. 
This corresponds to the echelle orders 24, 25, and 26 spread over the three detectors and the suborder on the first detector of echelle order 23.
{\tt viper} treats those detector chunks separately and combines the results afterwards as described by Eqs.~\eqref{eq_RV}~and~\eqref{eq_eRV}.
The wavelength region between 20\,600 and 21\,400~\r{A} suffers from a lack of gas cell lines (see Fig. \ref{fig:set_cell}), and the wavelength regions below 20\,600~\r{A} and above 24\,700~\r{A} are contaminated by strong telluric bands.
Tests have shown that for a spectral suborder of 2048 pixels, a polynomial degree of two for the wavelength solution and a symmetric Gaussian for the IP model deliver the best results for our CRIRES$^{+}$ observations.

\begin{table*}
\caption{Sample of RV standard stars. }.            
\label{tab_sources}     
\centering                          
\begin{tabular}{l c c c c c c c c c c}        
\hline\hline                 
Target & Sp.type & m$_K^{a}$ & $N_{\mathrm{HARPS}}^{b}$ & rms$_{\mathrm{HARPS}}^{b}$ & $N_{\mathrm{CR+}}^{c}$ & rms$_{\mathrm{CR+}}^{c}$ & rms$_{\mathrm{CR+}}^{c}$ & rms$_{\mathrm{CR+}}^{c}$ & time span$^{c}$ \\    
 & & & & & cell/no cell & ($v_\mathrm{star}$, cell) & ($v_\mathrm{star}$, no cell) & ($v_\mathrm{atm}$, cell) \\
  & & [mag] & & [m/s] &  & [m/s] & [m/s] & [m/s] \\    
\hline                        
GJ\,447 & M4.0$^{d}$ &  5.65 & 201 & 2.76 & 21 / 21 & 4.16 & 12.56 & 11.60 & Jun 2021 - Jun 2023\\
GJ\,588 & M2.5$^{e}$ &  4.76 & 270 & 1.74 & 61 / 37 & 3.01 & 9.89 & 7.21 & Jun 2021 - Aug 2023\\
GJ\,229A & M1.0$^{f}$ &  4.63 & 201 & 2.60 & 29 / 29 & 3.99 & 15.57 & 16.93 & Aug 2021 - Jan 2024\\
GJ\,784 & M0.0$^{g}$ &  4.28 & 40 & 5.76 & 76 / 46 & 5.01 & 10.89 & 7.35 &  Jun 2021 - Oct 2023\\
\hline                                   
\end{tabular}
\tablefoot{$^{(a)}$ \citet{cutri} $^{(b)}$ \citet{trifonov} $^{(c)}$ This work $^{(d)}$ \citet{t447} $^{(e)}$ \citet{t588} $^{(f)}$ \citet{t229} $^{(g)}$ \citet{t784}.}
\end{table*}

\begin{figure*}
\centering
\includegraphics[ width=0.49\textwidth]{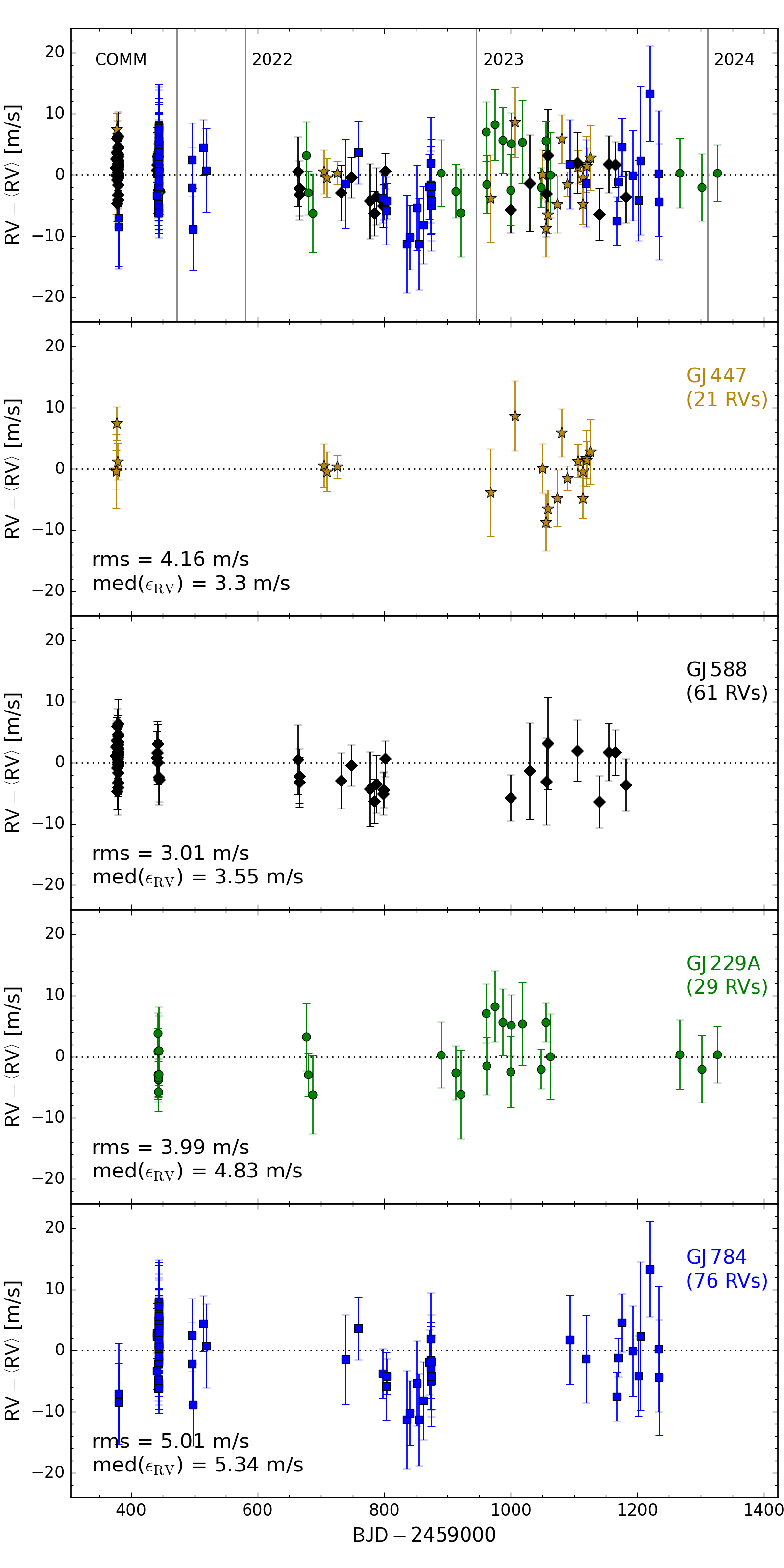}
\includegraphics[ width=0.49\textwidth]{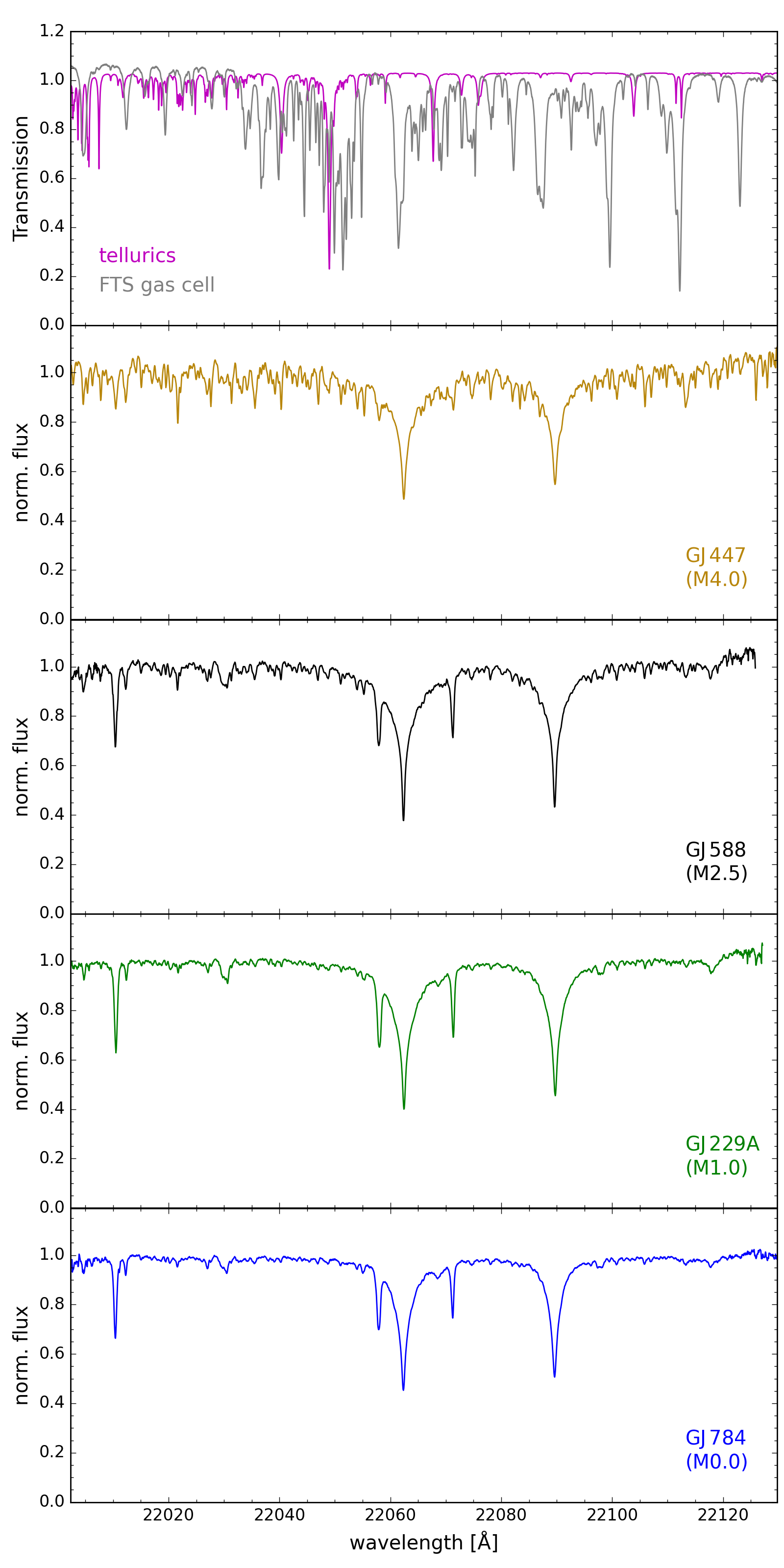}
\caption{{Left}: RV time series over 2.5 years for all four stars together (top) and for each source (bottom). Right: Gas cell spectrum obtained with the FTS and the telluric spectrum (top) and the telluric-free stellar templates (bottom) for one order created with {\tt viper}, as described in Sect.~\ref{sec:model_tell}.}
\label{fig:RV_all}
\end{figure*}

\begin{figure*}
\centering
\includegraphics[ width=0.49\textwidth]{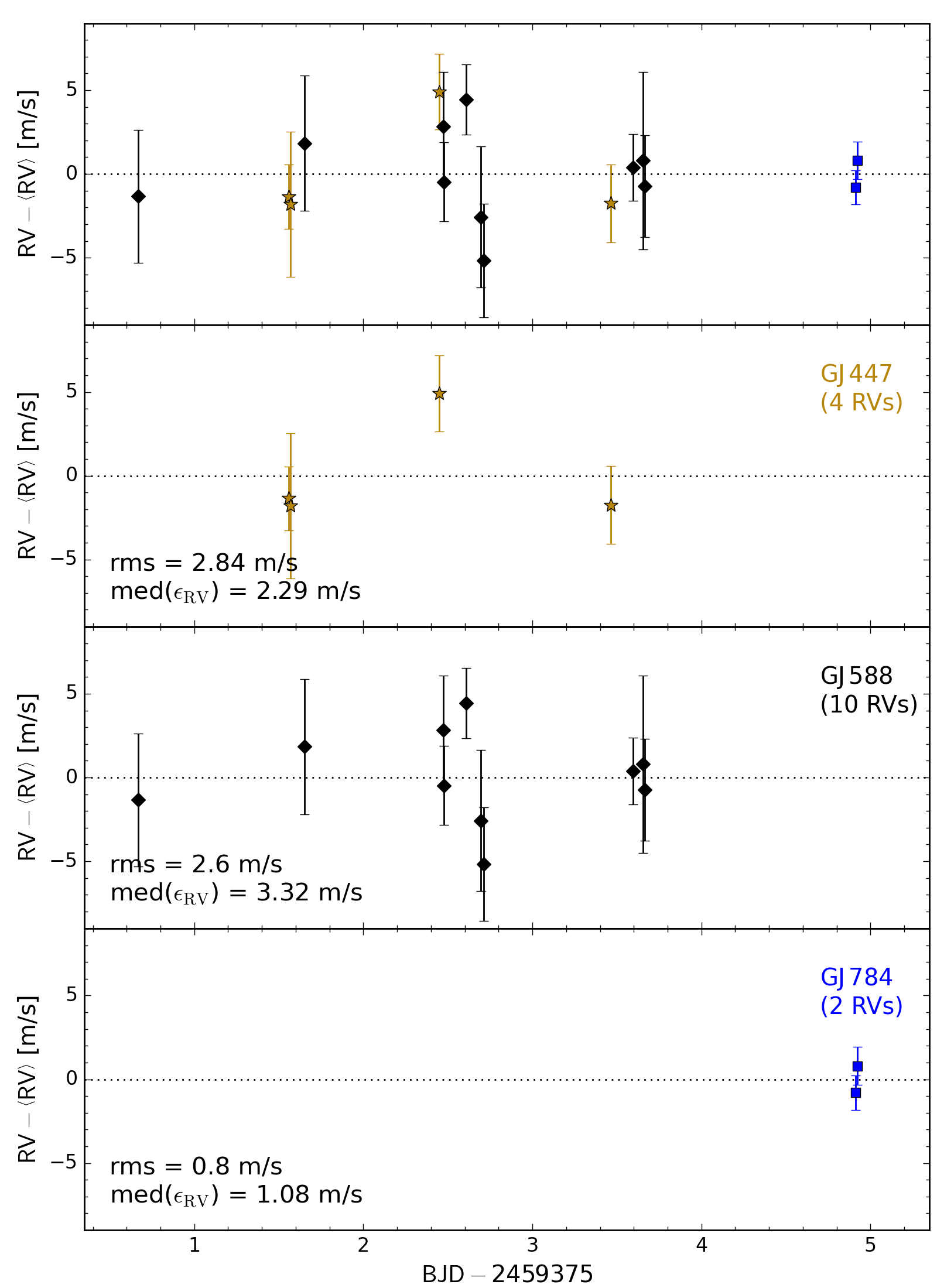}
\includegraphics[ width=0.49\textwidth]{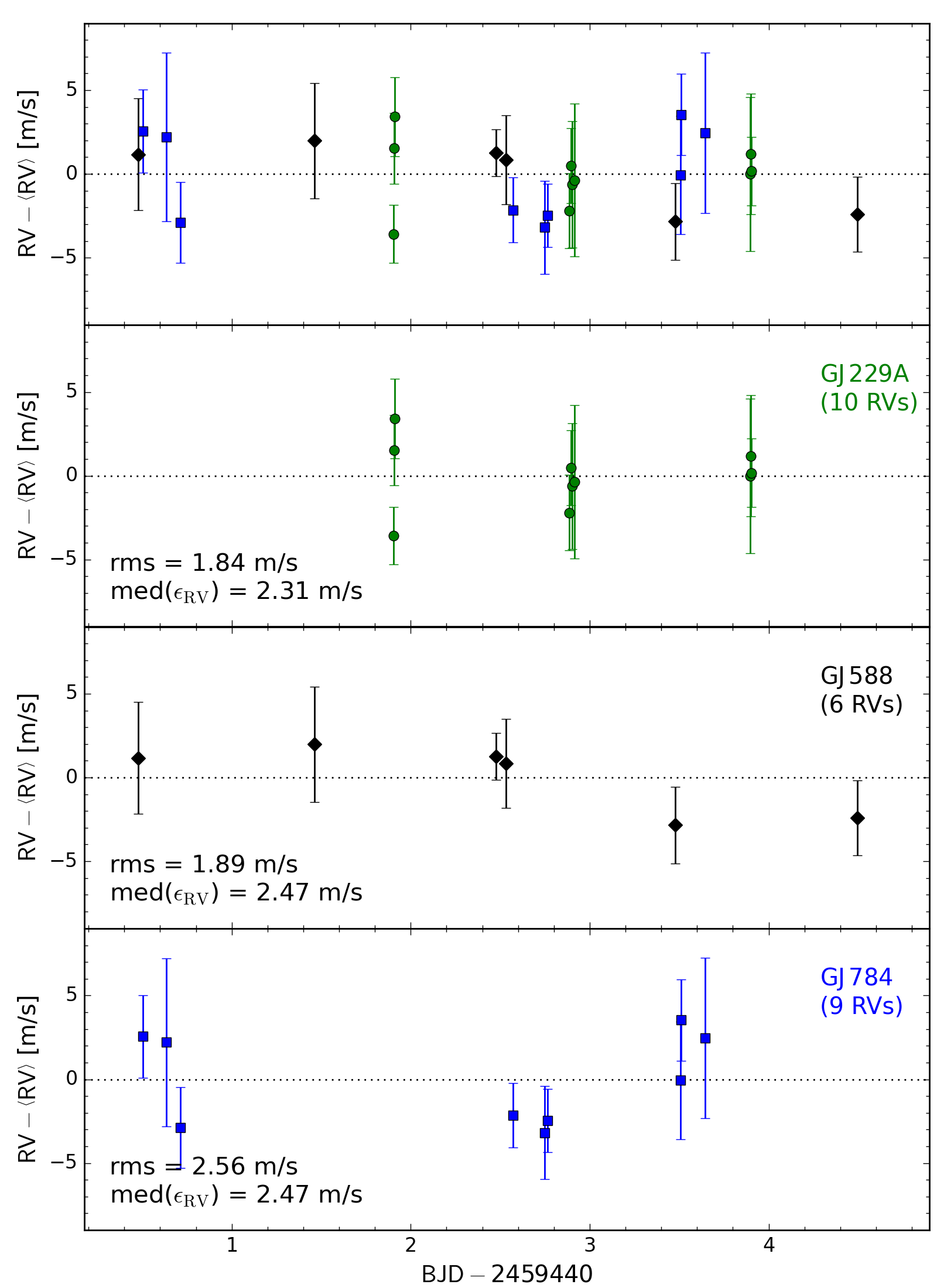}
\caption{Scatter of RV measurements for the standard star sample on a timescale of days. Left: COMM3 observations from June 2021. Right: COMM4 observations from August 2021. RV precision values of the order of 2-3~m/s were reached over a time range of several days. }
\label{fig:RV_comm}
\end{figure*}

\begin{figure*}
\centering
\includegraphics[ width=0.99\textwidth]{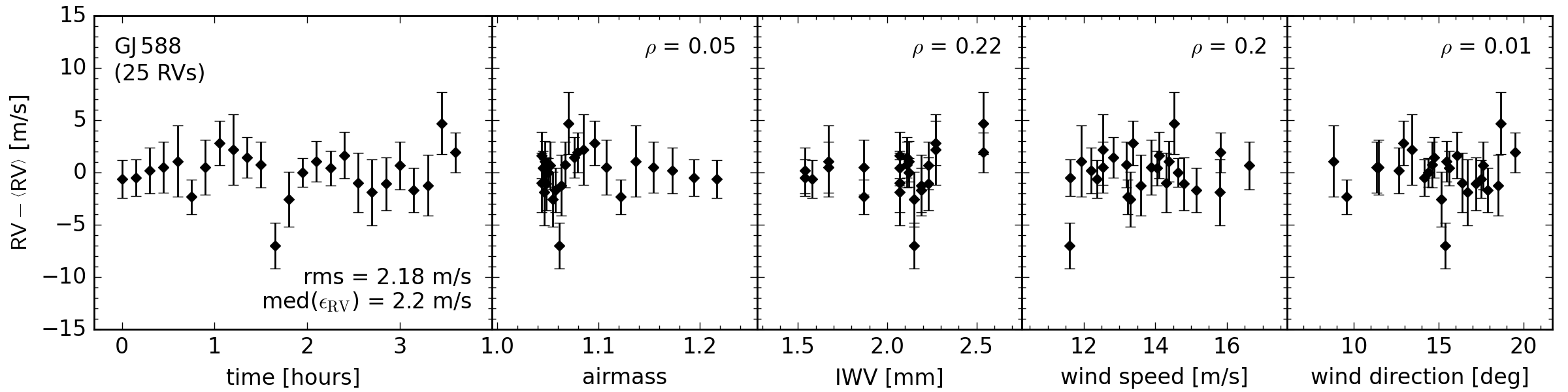}
\includegraphics[ width=0.99\textwidth]{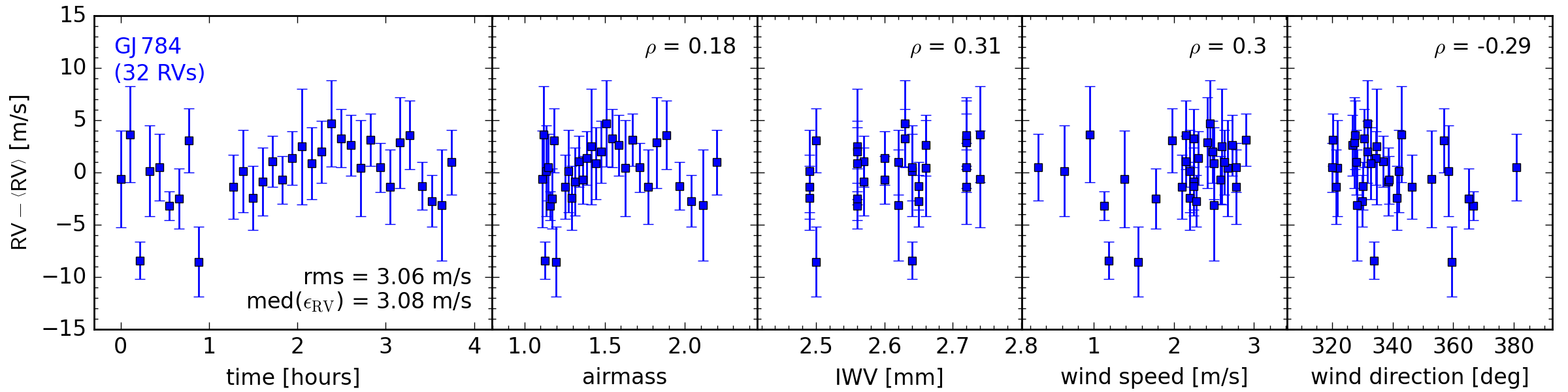}
\caption{Scatter of RV measurements for GJ\,588 (14 June 2021, top) and  GJ\,784 (17 August 2021, bottom). Left: Time series performed over a time range of around 3.7~hours. Right: Relations between the RV and various weather parameter together with the corresponding Pearson correlation coefficient, $\rho$. Even with changing weather parameters, an RV precision of 2.18~m/s for GJ\,588 and 3.06~m/s for GJ\,784 is reached.}
\label{fig:RV_TS}
\end{figure*}

\begin{figure*}
\centering
\includegraphics[ width=0.99\textwidth]{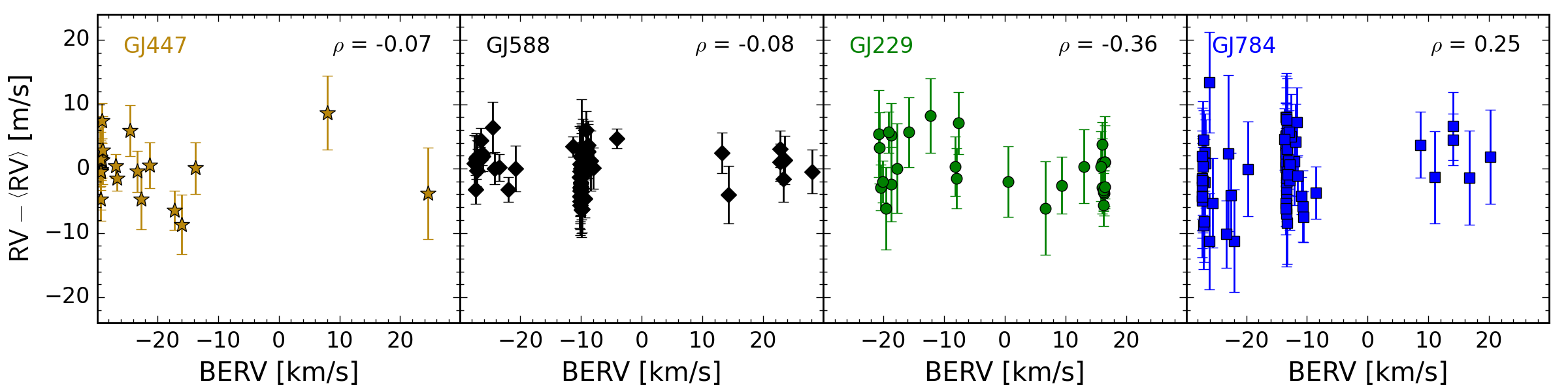}
\caption{RV measurements plotted against the BERV.}
\label{fig:RV_berv}
\end{figure*}

\section{Results}
\label{sec:results}

In this section we present the RV results obtained with {\tt viper} and the method of the telluric forward modelling. 
First, we present the RV results for the observations with the gas cell. 
Second, we study the RV results for observations performed without a gas cell, demonstrating that telluric lines can serve as wavelength reference alone. For this, we studied the telluric line stability on longer timescales.

\subsection{Template spectra}

A telluric-free stellar template was generated for each star as outlined in Sect. \ref{sec:model_tell}. Around ten observations (without the gas cell) at different epochs were combined to create an initial telluric-free template (see Fig.~\ref{fig:Tell_corr_star} for GJ\,588). To improve the quality of the template, another three iterations of template creation were performed, always using the previous template as reference input. Tests with more observations or iterations of template creations did not significantly improve the final template or obtained RV precision. 

The final template for each star, for a selected wavelength range, is plotted in the right panels of Fig.~\ref{fig:RV_all}. The stellar line density varies for the selected stars, depending on the spectral type (see \mbox{Table}~\ref{tab_sources}). Taking into account that the RV precision depends on the stellar line depth and density, it is expected to vary within our sample.

\subsection{RV precision using the gas cell}

Figure~\ref{fig:RV_all} (left) shows the 2.5 year RV time series for our standard star sample, the rms values are listed in \mbox{Table}~\ref{tab_sources} (rms$_{\mathrm{CR+}}$($v_\mathrm{star}$, cell)).
For ease of comparison, individual targets are shown with the same coloured symbol in each plot.
The results from the two commissioning runs are shown in Fig.~\ref{fig:RV_comm}, which gives an idea of the RV stability on a timescale of several days. Furthermore, during both commissioning runs, two time series of observations covering 3.7~hours were performed for GJ\,588 and GJ\,784 and are plotted in Fig.~\ref{fig:RV_TS}.

For the observations taken within a few hours or days, the barycentric Earth radial velocity (BERV) hardly changes and the impact of incorrect telluric modelling would be small, which also leads to a higher RV accuracy. The situation is different for observations taken over a period of one year. 
In Fig.~\ref{fig:RV_berv} the RVs are plotted against the BERV. We find no correlation between these two values, but detect a higher Pearson correlation coefficient for GJ784 and GJ229, which could be explained by the lower stellar content, compared to GJ558 and GJ447, resulting in a stronger influence of the telluric lines.

\begin{figure*}
\centering
\includegraphics[ width=0.99\textwidth]{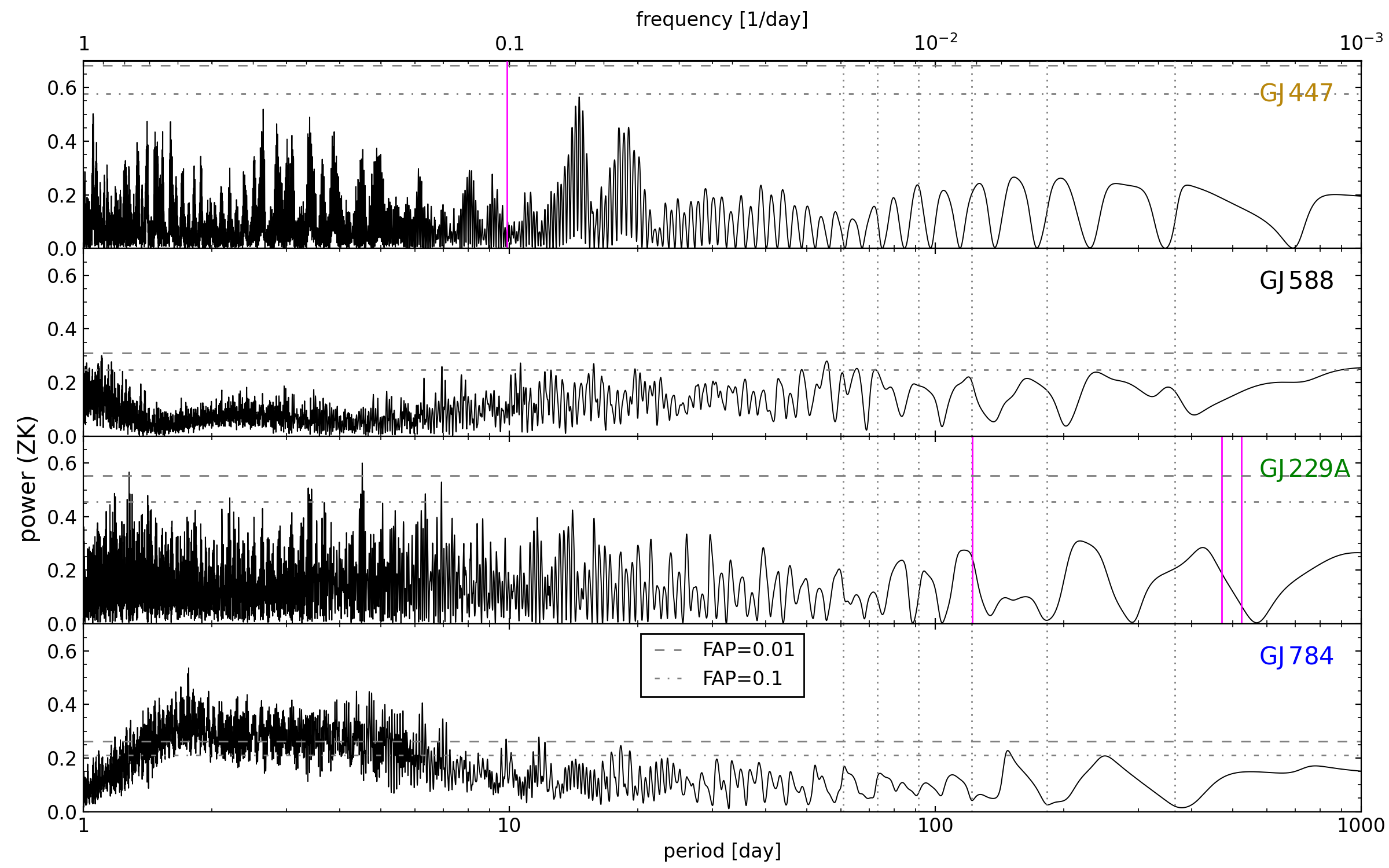}
\caption{Lomb-Scargle periodograms for the RV standard sample. The vertical dotted grey lines represent the periods of the harmonics of one year (from 365~days to 365/6~days). The vertical magenta lines denote known planet signals for GJ\,447 and proposed planet signals for GJ\,229A. The horizontal grey lines represent the false alarm probability level of 1\% (dashed line) and 10\% (dash-dotted line).}
\label{fig:gls}
\end{figure*}

A Lomb-Scargle periodogram \citep{gls} for each star, the false alarm probability \citep[][]{fap} levels, known planet signals, and the harmonics of the Earth's period of one year are shown in Fig.~\ref{fig:gls}. Simulations have demonstrated that wrong line profiles from the tellurics can lead to strong period signals at these harmonics \citep[e.g.][]{wang}. Similar issues are expected for other modelling errors. Nevertheless, we do not detect such a signal for any of our stars.

\subsubsection{GJ\,447 (Ross\,128)}
\label{sec:GJ447}

GJ\,447 is an M4.0V with m$_K$~=~5.65 mag, which has one confirmed planet with an orbital period of 9.87~days and a semi-amplitude of 1.39~m/s \citep{ross128}. 
This semi-amplitude is likely too small to be detectable with CRIRES$^{+}$, thereby allowing the target to be treated as an RV standard star. As the latest spectral type, GJ\,447 has the highest stellar line density of our sample.

During COMM3, four observations over three nights were taken, reaching a precision of 2.84~m/s (Fig.~\ref{fig:RV_comm}). On a timescale of 25 months, a precision of 4.16~m/s could be reached for 22 observations.  
We obtain a predicted precision limit of 1.7~m/s for GJ\,447 for an observation with an S/N of 260. This S/N corresponds to the mean S/N of all our observations of this target.
We calculated the predicted precision limit directly from the observation by applying the method described by \citet{butler} and \citet{bouchy}. In doing so, we considered all orders used for the RV calculation but masked pixels at the edges of the order. However, we did not mask regions that are contaminated by telluric lines.

As expected, the planetary signal cannot be detected in the Lomb-Scargle periodogram of the CRIRES$^{+}$ RV data (Fig.~\ref{fig:gls}). This is likely due to the low semi-amplitude of 1.39~m/s, as well as due to our time sampling, which is roughly one observation once every seven to ten days and therefore close to the planet's period.

\subsubsection{GJ\,588 (CD-40 9712)}
\label{sec:gj588}

GJ\,588 is an M2.5 star with m$_K$~=~4.76 mag. It is a high proper motion star with no confirmed planets. In our sample, it is the star with the lowest RV variability in the HARPS dataset ($\mathrm{rms}_\mathrm{HARPS}$~=~1.74~m/s).

On 14 June 2021 (COMM3), 25 measurements of the star were made within 3.75~hours (Fig.~\ref{fig:RV_TS}). Over the course of the observations the airmass ranged from 1.04 to 1.24. With an IWV at zenith rising from 1.5~mm to 2.6~mm, the telluric impact from water molecules became stronger towards the end. Wind speeds from 12-17~m/s at the ground were measured. 
Despite changing weather conditions, an rms of 2.2~m/s is reached 
and the correlation between the measured RVs and the weather parameters is weak with Pearson correlation coefficients below 0.2 (Fig.~\ref{fig:RV_TS} top).

GJ\,588 was observed another ten times during four nights in the same commissioning run. Here an rms of 2.6~m/s was achieved. Likewise, for the six observations during five observing nights in COMM4 an RV precision of 1.89~m/s could be obtained (Fig.~\ref{fig:RV_comm}).
For the total number of 61 observations performed over 26 months, an RV precision of 3~m/s is reached.
The average S/N value for all our observations of GJ\,588 is 315, and we calculate a predicted precision limit of 1.6~m/s for an observation with this S/N using the method described above.

\subsubsection{GJ\,229A (HD\,42581)}
\label{sec:gj229}

GJ\,229A is an M1.0 star with m$_K$~=~4.17 mag.
Until recently, it appeared to have a companion with an orbital period of 52\,890~days and amplitude of 17.06~m/s \citep{gj229}, which turned out to be a binary of two brown dwarfs orbiting each other \citep{xuan}.
Further, \citet{gj229} claimed that GJ\,229A has two planets with orbital periods of 523.2~days (${K_b=}$~1.63~m/s) and 122~days (${K_c=}$~2.15~m/s), while \citet{tuomi} claimed one planet with an orbital period of 471~days (${K_b=}$~3.83~m/s). Only recently, \citet{gj229_nop} published a paper showing that these signals are most likely caused by stellar activity. 
Due to the long period of the binary and the small amplitudes of the proposed planets, GJ\,229A is an usable RV standard star. 

The star was observed ten times during three consecutive nights in COMM4, leading to an rms of 1.84~m/s (Fig.~\ref{fig:RV_comm}). Over a time range of 29 months, an rms of 3.99~m/s is reached for 29 observations. 
The predicted precision limit for this target is 1.4~m/s for an observation with an S/N of 337, which corresponds to the average S/N of all observations of GJ\,229A. 

In recent years, a number of papers have been published on GJ\,229, showing that it is a complex system that is not yet fully understood. 
We included the proposed orbital periods of 523, 122 and 471~days in the Lomb-Scargle periodogram (Fig.~\ref{fig:gls}), but find no significant signal.

\begin{figure*}
\centering
\includegraphics[ width=0.49\textwidth]{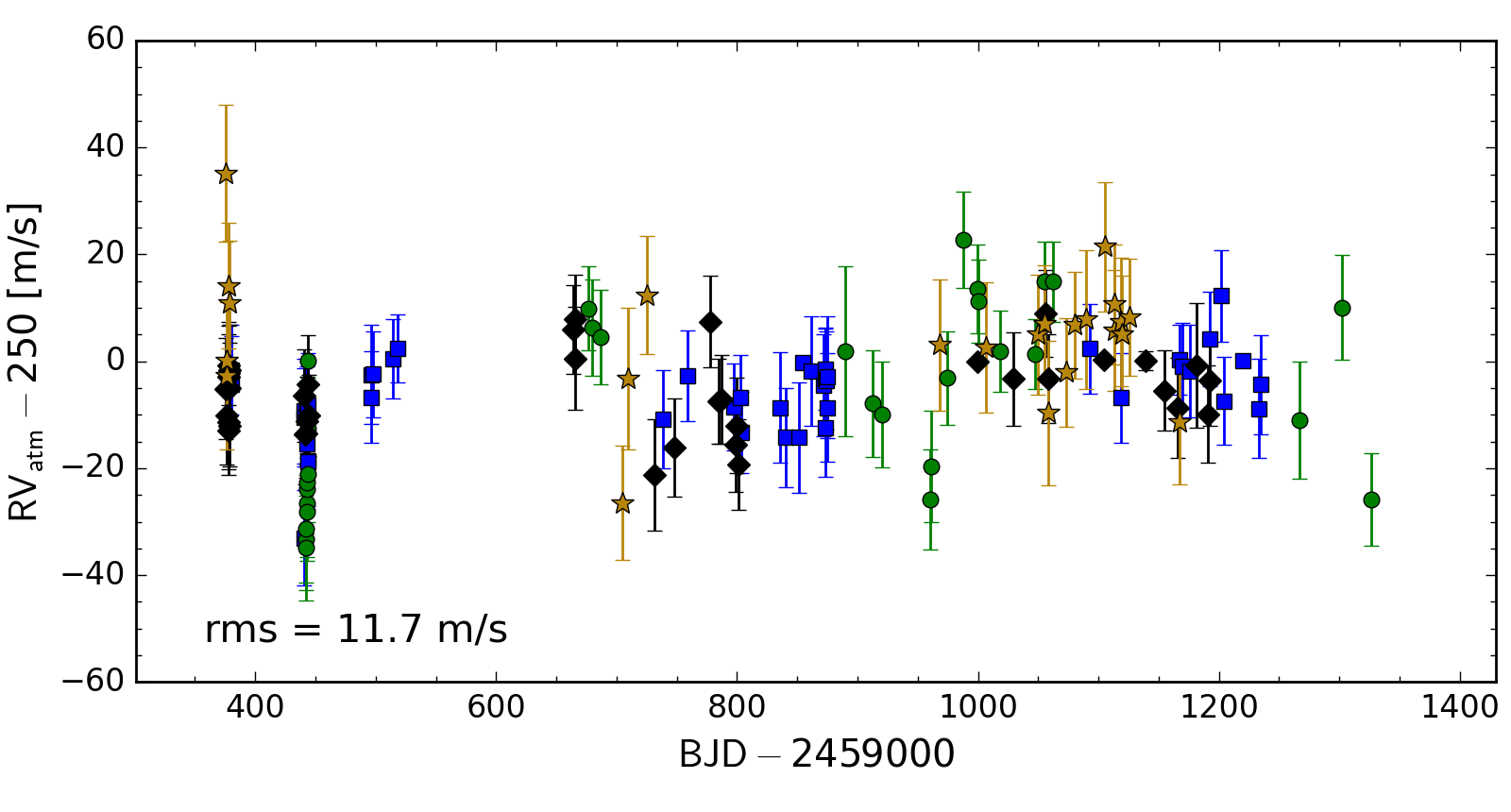}
\includegraphics[ width=0.49\textwidth]{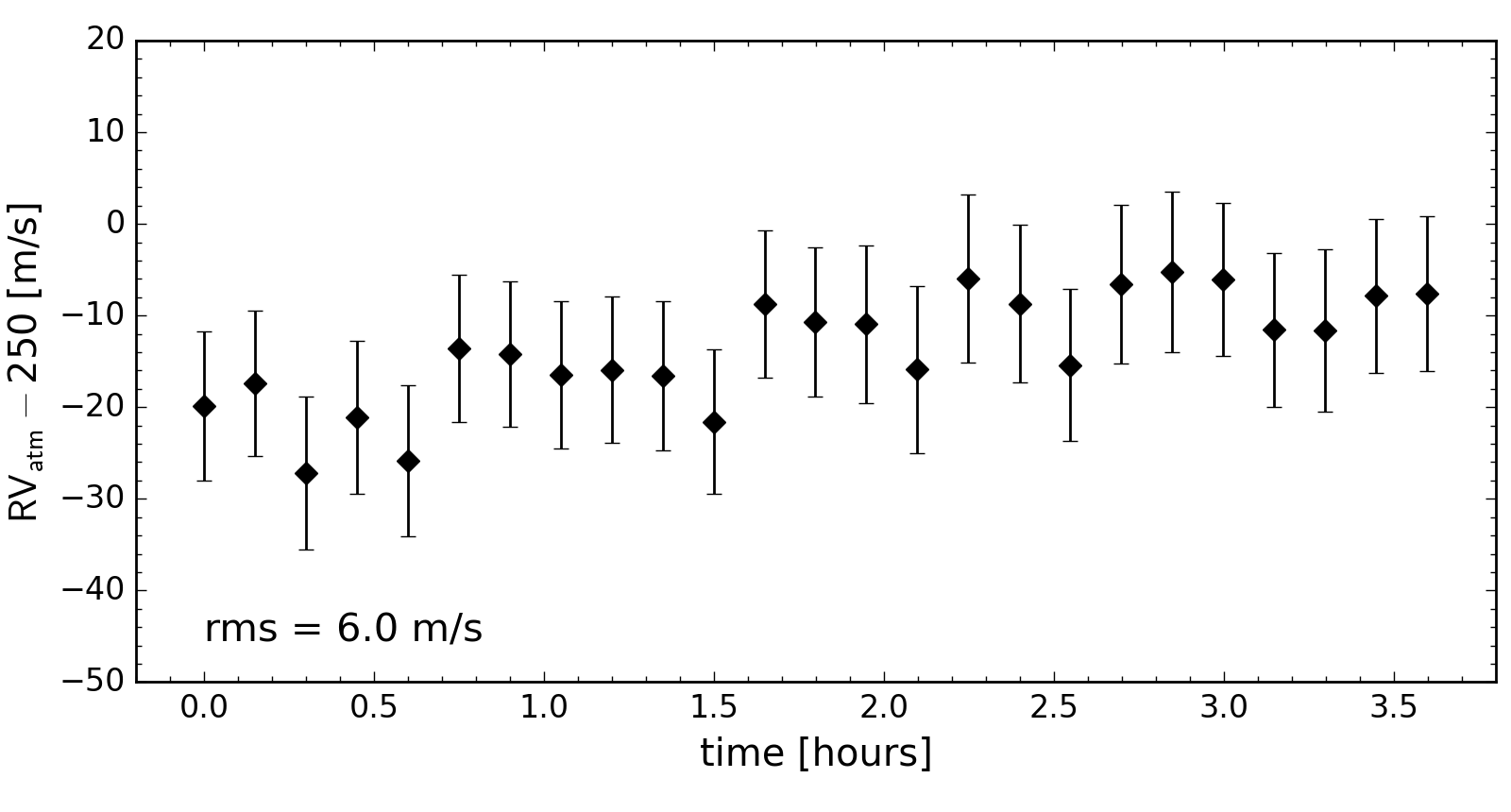}
\caption{Telluric Doppler shifts obtained for one spectral order (22730\,-\,22870~\r{A}), where the telluric lines dominate and nearly no stellar lines are present. Left: Telluric Doppler shifts as a function of time (same colour coding as in the previous plots). The rms of the 134 measurements is 11.7~m/s. {Right}: Telluric Doppler shift obtained from the time series observations of GJ\,588 from June 2021. The rms of the 25 measurements is 6~m/s over a time range of around 3.75~hours.}
\label{fig:tellshift}
\end{figure*}

\begin{figure}
\includegraphics[ width=0.49\textwidth]{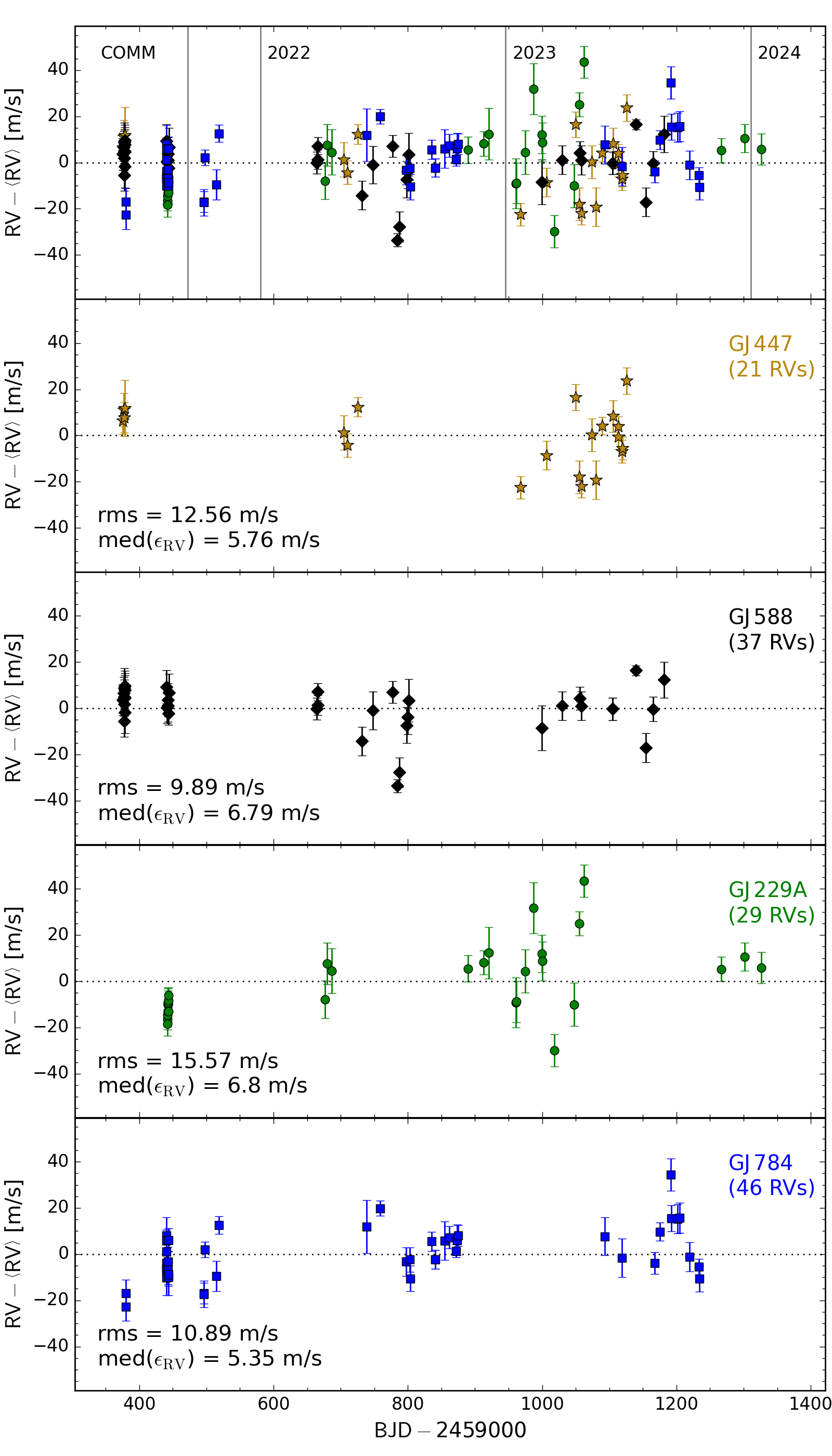}
\caption{RVs for the standard star sample observed without the gas cell as a function of time (colour coding as in previous plots). Telluric lines serve as the wavelength reference.}
\label{fig:RV_all_noSGC}
\end{figure}

\subsubsection{GJ\,784 (HD\,191849)}
\label{sec:gj784}

GJ\,784 is an M0.0 dwarf with m$_K$~=~4.28 mag. Though it has no confirmed planet, it is in our sample the star with the strongest RV variations in the optical data ($\mathrm{rms}_\mathrm{HARPS}$~=~5.76~m/s). 

Similar to GJ\,588, our observations can probe timescales of hours, days, and months. 
A time series observation over 3.7~hours was performed on the night of 17 August 2021 (COMM4). During the time series, the airmass of the star changed from 1.1 to 2.2. The IWV at zenith was nearly constant, with a mean value of 2.6~mm. This resulted in strong telluric lines, especially towards the end of the night.
The wind speed at the ground on this night was close to zero. Even with the impact of strong telluric lines, which are harder to model (see Sect. \ref{sec:model_tell}), an rms of 3~m/s is reached (Fig.~\ref{fig:RV_TS} bottom). This is slightly worse than what was obtained over a similar time range for GJ\,588. This shows the impact of the weather conditions as well as of the stellar line density. However, we cannot distinguish the contribution of the different factors. Again, the correlation between the RV trend and airmass or IWV is weak, with Pearson correlation coefficients below 0.3.

Besides the high cadence time series, GJ\,784 was observed for another nine times for three nights in COMM4, reaching a precision of 2.56~m/s (Fig.~\ref{fig:RV_comm}). During COMM3 just two observations within one night were taken (rms = 0.8~m/s). On the long timescale, GJ\,784 has an RV scatter of 5~m/s over 28 months. 
For all our observations of GJ\,784, we obtain an average S/N value of 314 and a corresponding predicted precision limit of 2.6~m/s.

\subsection{Stability of atmospheric lines}
\label{sec:tellshift}

As mentioned in Sect. \ref{sec:model_tell}, {\tt viper} allows the telluric Doppler shift to be a variable parameter, under the premise of existing cell lines.
The following results were obtained from the same observations, which already have been used to calculate the stellar RVs. This time, we were not interested in the stellar, but the telluric Doppler shifts, which  are a by-product of the modelling process in {\tt viper}. 
We selected one spectral order (22\,730--22\,870~\r{A}) with dominant tellurics and gas cell lines but hardly any stellar lines. 
For the remaining orders, we measured shifts with a larger scatter due to the blending of the stellar lines.

In the left panel of Fig.~\ref{fig:tellshift}, the telluric Doppler shifts are plotted for 134 observations. Over 31 months the telluric lines show an overall variation of 11.7~m/s around a mean offset of 250~m/s.
For the inter-night telluric shifts, we get a variation of 6~m/s from the 3.7 h time series observation of GJ\,588 (right panel of Fig.~\ref{fig:tellshift}).
All of this is in good agreement with the results reported by \citet{figueira}, who measured with HARPS a variation of 10~m/s around a zero offset of 222~m/s.
This agreement shows that the atmospheric lines in the NIR (mainly CH${_4}$ in the selected order) are as stable as the O$_2$ lines in the optical.
The difference of the zero-offset between our results and those of \citet{figueira} might be explained by different species, wavelength ranges, times, locations, and weather conditions.

The HITRAN database \citep{hitran}, which is used by the {\tt LBLRTM}, gives the line positions for a pressure of 1~atm. Wavelength shifts caused by pressure variations are not applied to the synthetic spectra that we used for {\tt viper}.

\subsection{RV precision using tellurics as the wavelength reference}
\label{sec:RVtell}

For most science studies, observations with a gas cell are used to obtain high-precision RVs. But there are some cases where it could be helpful to get RVs from observations taken without a cell. For instance, faint stars are normally observed without the cell to maximise the S/N. 
Further, such an approach would enable an RV study of observations of other science projects, where the contamination by cell lines is not desired (e.g. in the study of planetary atmospheres).

Since the CRIRES$^{+}$ spectrograph is not stabilised to a level needed for precise RV measurements, an improper wavelength correction would lead to instrumental drifts up to 1 km/s. Taking advantage of the telluric forward modelling as offered by {\tt viper}, the telluric lines can be used as reference. As shown above, these are stable down to around 10~m/s, which is a substantial improvement over traditional wavelength calibration.
{\tt viper} is one of the few RV codes that allow forward modelling using telluric lines alone as the wavelength calibration.

To assess the RV precision reachable with wavelength calibration based solely on the telluric lines, we re-analysed our sample of RV stars. Since on most nights observations were made with and without the cell, we have a sample of no-cell observations at hand. The modelling within {\tt viper} follows the same procedure as described above, just without the modelling of the cell lines. Furthermore, the wavelengths of the telluric lines are fixed (meaning no telluric Doppler shift is applied) as they serve as the wavelength reference.

The RV variations are plotted in Fig.~\ref{fig:RV_all_noSGC} and listed in \mbox{Table}~\ref{tab_sources} (rms$_{\mathrm{CR+}}$($v_\mathrm{star}$, no cell)), together with the variations of the telluric Doppler shifts (rms$_{\mathrm{CR+}}$($v_\mathrm{atm}$, cell)).
As mentioned before, the telluric Doppler shifts have been obtained from observations using the cell. In most cases, observations with and without the gas cell were taken within a few minutes of one another.

This time the RV precision varies between 10 and 16~m/s, or about two to three times worse compared to the RV precision of the gas cell observations. Nevertheless, with the telluric lines being stable down to 10~m/s, this is consistent with expectations.

Since the telluric lines are most stable, down to 7~m/s, for the observations of GJ\,588 and GJ\,784, we expect the RV precision to be the best for these targets, as is the case. 
The best RV precision of 9.89~m/s is obtained from the 37 observations of GJ\,588. We get a comparable RV precision of 10.89~m/s for the 46 observations of GJ\,784, even though this star has shown the largest RV scatter for the gas cell observations.
For the 21 observations of GJ\,447, we reach a precision of 12.56~m/s. The worst precision, with a value of 15.57~m/s, is obtained for the 29 observations of GJ\,229A. This can be explained by the strong variations of the telluric Doppler shifts during the observations of this star (see Fig.~\ref{fig:tellshift}).

\section{Known problems and outlook}

Although {\tt viper} performs well, we plan to further improve the code by addressing open issues.
One area would be a better modelling of the atmosphere lines and the IP. Even if the telluric modelling works well, especially for the telluric standard stars, we face problems for the telluric correction of some M~dwarfs observations. This can lead for example to strong residuals or an incorrect wavelength solution. Despite the fact that these are seen in only a few spectra, and most are averaged out when combining several spectra to produce the template, an improvement of the atmosphere model is under investigation. One aim is to include an option to model the changes of the telluric line profiles. Profile changes are expected to be introduced due to pressure variations or low-elevation observations, which can lead to asymmetric line profiles \citep{smette, allart}. 

Apart from this, a correct parameterisation of the IP used in the modelling is essential \citep{endl, hatzes}.
Asymmetries of the IP profile can create RV shifts that are not real, but falsely can be interpreted as a planetary signal. {\tt viper} already offers a number of different IP profiles, but tests are still ongoing to improve the results and maybe come up with further methods.

Whereas the telluric forward modelling was mainly used and tested for observations in the NIR, the idea is to extend this method to shorter wavelengths, down to the optical region. Here this technique would enable us to use parts of the spectrum not covered by iodine cell lines. For instruments operating in the optical, just half of the observed spectral range is covered by the iodine cell spectrum, leaving the rest unused.

Another extension regarding usable wavelength ranges has to be done for the CRIRES$^{+}$ instrument itself. The results presented here were using observations in the \textit{K}~band. Nevertheless, as the gas cell also covers the \textit{H}~band, and telluric lines are present all over the entire NIR range, the next step would be to investigate the RV precision apart from the \textit{K}~band. 
Finally, as a long-term goal we wish to extend the {\tt viper} support to more spectrographs.

\section{Summary}

We have demonstrated that it is possible to apply a simple telluric forward model to reach an RV precision down to 3~m/s with CRIRES$^{+}$ in the \textit{K}~band over timescales of 2.5 years for bright M~dwarfs using a gas cell.
While most telluric correction codes require modelling of many layers of the Earth's atmosphere, {\tt viper} works with synthetic spectra and scaling exponents, plus a parameter for telluric Doppler shifts. 
Thanks to this simple model, {\tt viper} needs only a few seconds of processing time and is therefore faster than other available telluric correction codes. 
Apart from that, all processing -- from the creation of a telluric-free stellar template to the determination of RV values -- can be done without any additional correction codes.

Using observations from CRIRES$^{+}$, we have demonstrated that {\tt viper} is able to successfully model telluric lines in the NIR. For the example of a telluric standard star, we obtain rms values between 0.6\% and 1\% for the corrected spectrum depending on the strength of telluric lines. Nevertheless, it should be mentioned that {\tt viper}, just like any other correction code, is not able to model saturated lines correctly. With observations distributed over a longer time period, residuals from the correction can be reduced or even be removed via the co-adding of several spectra, as shown in the example of the M~dwarf GJ\,588.

We used {\tt viper} on a number of RV standard stars observed with CRIRES$^{+}$ over a time span of more than 2.5 years. An RV precision of around 2~m/s was reached on short to medium timescales of hours, days, and weeks. The best long-term RV precision of 3~m/s was reached for the M2.5 star GJ\,588 over a time range of 26 months. In our sample, GJ\,588  is the star with the lowest RV variation in the optical as well. For the rest of the sample, precision values between 4~m/s and 5~m/s were obtained on long timescales.
This shows that CRIRES$^{+}$ is a suitable instrument for high-precision RV measurements in the NIR.  {\tt viper} is the officially recommended RV pipeline for CRIRES$^{+}$ data, and the user has an easy and fast tool at hand to process the data. The code comes with a manual and demo data.

In addition, we have studied the stability of telluric lines in the NIR, confirming the variation of around 10~m/s already reported in other studies, including \citet{figueira}. Furthermore, we calculated RV values for observations performed without the use of a gas cell. Applying the forward modelling using only telluric lines as the wavelength reference, we were able to obtain an RV precision down to 10~m/s on a timescale of more than two years, which is in full agreement with the stability of the telluric lines.

In this paper only results from CRIRES$^{+}$ have been presented. Nevertheless, {\tt viper}  produces good results for other instruments as well \citep[][]{tls1, tls2, tls3, oes}.

\begin{acknowledgements}
Based on observations collected at the European Organisation for Astronomical Research in the Southern Hemisphere under ESO programmes 60.A.-9051, 108.22CH.001, 109.23FA.001, 110.2447.001, 111.24SC.002, and 112.25VY.002. 
CRIRES$^{+}$ is an ESO upgrade project carried out by Th\"uringer Landessternwarte Tautenburg, Georg-August Universit\"at G\"ottingen, and Uppsala University. The project is founded by the Federal Ministry of Education and Research (Germany) through Grants 05A11MG3, 05A14MG14, 05A17MG2 and the Knut and Alice Wallenberg Foundation. 
J.K. and A.H. acknowledge financial support from ESO and grant HA 3279/15-1 from the  Deutsche Forschungsgemeinschaft (DFG).
O.K. acknowledge support by the Swedish Research Council (projects 2019-03548 and 2023-03667).
We thank all colleagues involved in the development, installation and commissioning of CRIRES$^{+}$. Further, we thank observers and staff at the VLT providing us with observations and support. Acknowledgements also go to current users and testers of {\tt viper}.

\end{acknowledgements}

\bibpunct{(}{)}{;}{a}{}{,} 

\bibliographystyle{aa} 
\bibliography{aa53919-25}

\end{document}